\documentclass{article}

\usepackage{iclr2027_conference,times}

\usepackage{amsmath,amsfonts,bm}

\def\eqref#1{equation~\ref{#1}}

\def\1{\bm{1}}

\DeclareMathAlphabet{\mathsfit}{\encodingdefault}{\sfdefault}{m}{sl}
\SetMathAlphabet{\mathsfit}{bold}{\encodingdefault}{\sfdefault}{bx}{n}

\usepackage[T1]{fontenc}
\usepackage[utf8]{inputenc}
\usepackage{microtype}
\usepackage{booktabs}
\usepackage{tabularx}
\usepackage{array}
\usepackage{amsmath}
\usepackage{xcolor}
\usepackage{hyperref}
\usepackage{graphicx}
\usepackage{pgfplots}
\pgfplotsset{compat=1.18}
\usepackage{xurl}
\usepackage{verbatim}
\usepackage{xspace}
\usepackage{pifont}
\usepackage{float}
\usepackage{placeins}
\usepackage[font=small,labelfont=bf,subrefformat=parens,labelformat=parens]{subcaption}

\graphicspath{{figures/}}

\newcommand{\benchmarkname}{GameEngineBench\xspace}
\newcommand{\numtasks}{110}
\newcommand{\numrepos}{nine}
\newcommand{\numconfigs}{\geNumConfigsWord}
\newcommand{\bestpass}{\geBestPass}
\newcommand{\numunsolved}{\geNumUnsolved}

\newcommand{\cmark}{\ding{51}}
\newcommand{\xmark}{\ding{55}}
\newif\ifrevisionreview \revisionreviewfalse
\ifrevisionreview\usepackage[commandnameprefix=ifneeded]{changes}\else\usepackage[final,commandnameprefix=ifneeded]{changes}\fi

\newcommand{\geMutApplicable}{567}  %
\newcommand{\geMutKilled}{550}
\newcommand{\geMutKillRate}{97.0\%}

\newcommand{\geMutSurvivors}{17}
\newcommand{\geMutDistN}{67}

\newcommand{\geMutDistMissedByUI}{55}
\newcommand{\geMutDistClientOnly}{51}

\newcommand{\geSWPassTotal}{348}  %
\newcommand{\geSWPassCWFail}{64}  %
\newcommand{\geSWPassCWFailPct}{18.4\%}  %

\newcommand{\geNumClientWorldTasks}{53}
\newcommand{\geNumClientWorldTests}{119}
\newcommand{\geNumRankChangesWithoutCW}{3}  %
\newcommand{\geTagKappaTier}{0.972}
\newcommand{\geTagKappaWorld}{0.934}

\newcommand{\geTagNumTests}{1{,}114}
\newcommand{\geTagDisagreements}{34}

\newcommand{\geStaticOnlyRejections}{5}

\newcommand{\geNumConfigsWord}{eleven}
\newcommand{\geNumCells}{1{,}210}

\newcommand{\geBestPass}{46.4\%}

\newcommand{\gePassFableMax}{46.4\%}
\newcommand{\gePassCIFableMax}{[37.3, 55.5]\%}
\newcommand{\gePassCountFableMax}{51/110}

\newcommand{\gePassOpusFourEightMax}{31.8\%}

\newcommand{\gePassGptXhigh}{33.6\%}

\newcommand{\gePassGptHigh}{34.5\%}

\newcommand{\geNumUnsolved}{43}  %
\newcommand{\geNumSolvedByAny}{67}
\newcommand{\geNumSingleSolver}{16}

\newcommand{\geNumCompileFail}{24}

\newcommand{\geNumPairsDiffer}{27}
\newcommand{\geNumPairsTie}{28}
\newcommand{\geNumPairs}{55}

\newcommand{\geRefOfficialPass}{220/220}

\newcommand{\gePerTestFlips}{10}
\newcommand{\gePerTestCompared}{11{,}307}

\newcommand{\geEtoEOnlyFails}{209}
\newcommand{\geEtoEOnlyWouldBe}{493}
\newcommand{\geEtoEOnlyPct}{42.4\%}
\newcommand{\geEtoEOnlyIntFails}{69}
\newcommand{\geEtoEOnlyIntWouldBe}{269}
\newcommand{\geEtoEOnlyIntPct}{25.7\%}
\newcommand{\geEtoEOnlyDeltaMin}{12.7}
\newcommand{\geEtoEOnlyDeltaMax}{20.9}

\newcommand{\geNumTasksWithClients}{71}
\newcommand{\geNumTasksNoClients}{39}
\newcommand{\geCWDeltaMin}{3.6}
\newcommand{\geCWDeltaMax}{7.3}
\newcommand{\gePropFailNonLocal}{615}       %
\newcommand{\gePropFailNonLocalPct}{74.9}   %
\newcommand{\gePropFailOrdering}{316}       %
\newcommand{\gePropFailDistributed}{280}    %
\newcommand{\gePropFailCrossFile}{227}      %
\newcommand{\gePropFailLocalLogic}{618}     %

\newcommand{\geMutDistKilled}{65}

\newcommand{\geCovNumTests}{1{,}111}
\newcommand{\geCovTestsLocal}{673}
\newcommand{\geCovTestsCrossFile}{143}
\newcommand{\geCovTestsOrdering}{167}
\newcommand{\geCovTestsDistributed}{128}
\newcommand{\geCovTasksLocal}{106}
\newcommand{\geCovTasksCrossFile}{60}
\newcommand{\geCovTasksOrdering}{63}
\newcommand{\geCovTasksDistributed}{53}
\newcommand{\geCovTasksNoNonLocal}{14}
\newcommand{\geCovTasksOneNonLocal}{33}
\newcommand{\geCovTasksTwoNonLocal}{46}
\newcommand{\geCovTasksThreeNonLocal}{17}
\newcommand{\geCovTasksAnyNonLocal}{96}
\newcommand{\geCovTasksTwoPlusNonLocal}{63}

\newcommand{\geNumTestsUnit}{253}
\newcommand{\geNumTestsIntegration}{498}
\newcommand{\geNumTestsEtoE}{362}
\newcommand{\geNumTestsStatic}{one}
\newcommand{\geNumTestsMapped}{1{,}032}

\newcommand{\geMedianSolutionLines}{389}
\newcommand{\geMedianSolutionFiles}{two}
\newcommand{\geNumRenderingTasks}{six}
\newcommand{\geTimeLimitSec}{3{,}600}

\newcommand{\geNumWebAttempts}{274}
\newcommand{\geNumTimingSensitiveTests}{four}

\newcommand{\geTestFailCells}{821}
\newcommand{\geTestFailAttributable}{819}

\newcommand{\geNumSolvesWithTrajectory}{1{,}202}

\newcommand{\geNumSolvesNoTrajectory}{eight}

\newcommand{\geNumSingleSolverConfigs}{five}

\newcommand{\geNumBuiltTestFail}{821}

\newcommand{\geCompileFailClaudeCode}{4/550}
\newcommand{\geCompileFailCodex}{4/330}
\newcommand{\geCompileFailAntigravity}{3/110}
\newcommand{\geCompileFailQwenCode}{13/220}

\newcommand{\geCostPerPassFableMax}{42.20}

\newcommand{\geCostPerPassGptHigh}{7.01}

\newcommand{\geMutNumTasks}{72}
\newcommand{\geMutOtherN}{16}

\newcommand{\geMutLocalKillRateIntegration}{50.7\%}
\newcommand{\geMutLocalKillRateEtoE}{42.0\%}

\newcommand{\geExMatchLoopTests}{10}
\newcommand{\geExMatchLoopNearMiss}{Five}
\newcommand{\geExMatchLoopAuthorityFails}{six}

\title{\benchmarkname: Full-Session Verification of Coding Agents in a Networked Game Engine}

\expandafter\ifx\csname iflatexml\endcsname\relax\expandafter\let\csname iflatexml\expandafter\endcsname\csname iffalse\endcsname\fi
\iflatexml
\author{%
Brian La\textsuperscript{1}, Sungbin Jo\textsuperscript{1}, Sejoon Chang\textsuperscript{1},\\
Ben Kim\textsuperscript{1}, Junyoung Bae\textsuperscript{2}, Aamish Ahmad Beg\textsuperscript{3},\\
Sei Chang\textsuperscript{4}, Gonzalo Gonzalez-Pumariega\textsuperscript{5}\\
\textsuperscript{1}Nitrode, \textsuperscript{2}Nexon Intelligence Labs, \textsuperscript{3}Dartmouth College,\\
\textsuperscript{4}Columbia University, \textsuperscript{5}Cornell University
}
\else
\author{%
Brian La\textsuperscript{1} \And
Sungbin Jo\textsuperscript{1} \And
Sejoon Chang\textsuperscript{1} \And
Ben Kim\textsuperscript{1} \AND
Junyoung Bae\textsuperscript{2} \And
Aamish Ahmad Beg\textsuperscript{3} \And
Sei Chang\textsuperscript{4} \And
Gonzalo Gonzalez-Pumariega\textsuperscript{5} \AND
\makebox[\dimexpr\textwidth-2\tabcolsep\relax][c]{\textnormal{\begin{tabular}[t]{@{}c@{}}
\textsuperscript{1}Nitrode \quad \textsuperscript{2}Nexon Intelligence Labs \quad \textsuperscript{3}Dartmouth College \\
\textsuperscript{4}Columbia University \quad \textsuperscript{5}Cornell University
\end{tabular}}}
}
\fi

\iclrfinalcopy
\DeclareMathSizes{7.5}{7}{5}{5}
\begin{document}
\maketitle
\lhead{Preprint}

\begin{abstract}
Coding agents face new challenges in game development, where they must produce code that is not only syntactically correct but also consistent with engine-managed lifecycle, state, and networking behavior. Existing game-development benchmarks, however, do not combine assertions over engine-based runtime states with cross-client consistency, failing to evaluate how well generated code adheres to the desired behavioral specifications. To address these limitations, we present \benchmarkname, a benchmark of \numtasks{} C++ implementation tasks derived from \numrepos{} existing Unreal Engine 5 projects. Each task requires an agent to implement missing behavior within an established gameplay architecture. We also introduce \textsc{Full-Session Verification}, a testing framework that injects hidden tests after the agent finishes and evaluates the resulting implementation inside Play-in-Editor (PIE) sessions, with networked client worlds in \geNumTasksWithClients{} tasks. The tests evaluate Unreal's game loop, actor lifecycle, subsystems, and replication stack while directly inspecting runtime state. In \geNumClientWorldTasks{} tasks, \geNumClientWorldTests{} tests additionally inspect a distinct client world to determine whether the authoritative server state is correctly replicated. To our knowledge, this is the first coding benchmark to evaluate cross-client replication. Across \numconfigs{} model--harness configurations, the best-scoring configuration achieves \bestpass{} pass@1, while \numunsolved{} tasks remain unsolved by every evaluated configuration. These results demonstrate that compilation and shallow verification overestimate agents' ability to implement behavior that remains correct during full engine execution.

\end{abstract}

\section{Introduction}

Advances in coding agents have expanded their use beyond traditional software engineering tasks, leading to their increasing evaluation in complex game development settings \citep{chi2026gamedevbench,gamecraft2026,jamer2026,worldcoder2026,yin2026autoue}. An engine such as Unreal Engine 5 (UE5) provides the framework that determines when, where, and how gameplay logic is executed. Various components, including actor lifecycle (how the engine creates, initializes, updates, and destroys game-world objects, known as \textit{actors}), engine subsystems (engine-managed service objects), networking, rendering, UI, and scripts, must all be integrated for full correctness. For instance, a single feature may need to execute on the authoritative server (whose state is definitive), replicate to clients, update only for the local player, and clean up actors when destroyed or reused. The same feature may also have a visual component involving meshes, animations, particle effects, and heads-up displays (HUDs). Although prior work has measured limitations in models' visual reasoning capabilities \citep{chi2026gamedevbench}, whether their generated code behaves correctly inside a running game engine remains underexplored.

Most software engineering benchmarks evaluate features that are largely self-contained, allowing tasks to be independently verified. However, when multiple independent subsystems interact at runtime, coding agents must also satisfy correctness requirements that cannot be verified from the source code alone. We refer to behaviors that can only be verified at runtime as \textit{non-local properties} and divide them into two categories: temporal and distributed. A temporal property concerns the order and timing of events within the engine's lifecycle. Temporal violations include querying a subsystem before the engine has been initialized or invoking a callback after its target has been destroyed. Although ordering constraints are well established in the program verification literature \citep{strom1986typestate, ball2001temporal, meier2019lifestate}, coding benchmarks typically address them indirectly through input--output predictions \citep{gu2024cruxeval}. Lifecycle misuse is stated as a major cause of Android crashes \citep{fan2018android}, while ordering violations constitute a major class of concurrency bugs \citep{lu2008mistakes}. By contrast, a distributed property defines how the program state is maintained across separate runtime endpoints. In server--client applications, logic and state must remain consistent as updates are exchanged. Systems research has tested such distributed consistency through formal specifications of permitted behavior \citep{lamport2002tla, cheng2026specula}; however, coding-agent benchmarks generally do not test application-level replication across multiple running endpoints. 

For both types of non-local properties, code may still compile and pass isolated unit tests while still failing during full-system execution. These violations cannot be detected reliably through source inspection or unit testing alone, leading to critical errors in production. Although prior distributed systems research provides methods for testing properties such as database isolation \citep{kingsbury2020elle}, applying comparable verification to gameplay logic requires engine-aware test oracles and maintained behavioral specifications \citep{lamport2002tla}. Hence, game engines provide a useful evaluation setting, providing large, stateful systems with numerous non-local invariants where violations can be observed directly with the appropriate runtime verification methods.

We present \benchmarkname, a benchmark of \numtasks{} C++ implementation tasks embedded in \numrepos{} UE5 projects. For each task, the agent begins from a buildable project and must implement based on a task specification that describes the required behavior. To verify correctness, we introduce \textsc{Full-Session Verification}. After the agent completes its implementation, hidden tests are injected then executed in a Play-in-Editor (PIE) listen-server session.\footnote{PIE (Play-in-Editor) is Unreal's built-in mode for running the game. In a listen-server PIE session, one game instance acts as the authoritative server and hosts a single player. In \geNumTasksWithClients{} of the \numtasks{} tasks, however, further client instances connect to it, each maintaining their own copy of the game state.} PIE sessions run the entire game loop (repeating per-frame updates), including actor lifecycle, player controllers (player-control objects), and networking, allowing direct inspection of engine-managed state and detection of runtime failures.

We evaluate \benchmarkname on \numconfigs{} agent configurations. In each configuration, a model is paired with an \emph{agent harness}, software that enables the model to run commands and edit files. In this work, we highlight two main contributions:
\begin{itemize}
    \item \benchmarkname, a benchmark consisting of \numtasks{} C++ tasks drawn from \numrepos{} UE5 projects, each requiring proper integration with a running game engine.
    \item \textsc{Full-Session Verification} (FSV), a methodology for evaluating agent-generated code through controlled tests executed within a full game session, including assertions on networked client worlds when required by a task. To our knowledge, this is the first coding benchmark to offer cross-client replication testing.
\end{itemize}

We find that the best-scoring configuration solves \bestpass{} of the tasks, while \numunsolved{} tasks remain unsolved by every evaluated configuration. Our results highlight the importance of FSV in detecting failures that conventional single-world testing misses, as we observe that \geSWPassCWFail{} submissions that pass all single-world assertions fail tests on client worlds. Complementary mutant analysis further demonstrates the importance of our comprehensive, deterministic verification pipeline for evaluating agentic coding capabilities in modern game development. 

\section{Related Work}

\textbf{Software Engineering Benchmarks}. Evaluating coding agents on realistic software engineering tasks has become an active area of research. Most notably, SWE-bench evaluates whether models can resolve real GitHub issues in existing software repositories \citep{jimenez2023swebench}, while Terminal-Bench evaluates agents on realistic command-line tasks that require operating inside an execution environment \citep{merrill2026terminalbench}. A more recent benchmark, DeepSWE, handles long-horizon tasks with behavioral verifiers rather than inherited pull-request tests \citep{huang2026deepswe}. While these benchmarks provide executable feedback, they do not specifically evaluate lifecycle-dependent or distributed behavior. AppForge evaluates applications in an Android emulator \citep{ran2025appforge}, while SWE-Lancer and ABC-Bench interact with applications through a browser or HTTP client \citep{miserendino2025swelancer,abcbench2026}. In contrast, \benchmarkname uses each game project as both the source repository and execution environment, allowing tests to inspect behavior managed by the engine during a complete gameplay session.

\textbf{Game Development Benchmarks}. GameDevBench establishes game development as a setting for evaluating agents on multimodal tasks involving visual scenes, shaders, sprites, and animations \citep{chi2026gamedevbench}. AutoUE measures model capability in generating complete 3D environments within UE5 \citep{yin2026autoue}, while GameCraft-Bench, WorldCoder-Bench, and JAMER all evaluate game-development capabilities in other execution settings \citep{gamecraft2026,worldcoder2026,jamer2026}. Their verification methods, however, observe different aspects of correctness. GameDevBench primarily checks scene structure and callbacks rather than behavior under networked execution. GameCraft-Bench executes Godot games but judges gameplay recordings with a multimodal model. AutoUE generates complete UE5 games and evaluates them with LLM-generated play-testing commands run in the engine. WorldCoder-Bench performs runtime tests in a browser, but the generated program is responsible for exposing the state. JAMER checks for integrity, compilation, and crashes while evaluating behavioral correctness through similarity to a reference. PlayCoder evaluates generated GUI applications, including games, by running unit tests and then deciding behavioral correctness with PlayTester, an LLM agent that plays the application through its interface \citep{peng2026playcoder}. In contrast to prior benchmarks, \benchmarkname directly inspects engine-managed state during execution inside existing production C++ codebases, combining runtime assertions with verification across separate networked server and client worlds for the first time. Table~\ref{tab:benchmark-comparison} compares these benchmarks on four properties: whether the behavioral pass/fail label comes from executed tests rather than a model judge, whether the game loop runs, whether tests read state held by the running engine, and whether they check a separate client world of a networked session.

\textbf{Verification Work}. Reliable verification is necessary to minimize false positives and false negatives in measuring coding agent performance \citep{zhu2025abc}. Related verifier-side work on browser games injects runtime instrumentation and checks behavioral predicates \citep{jia2026gamegenverifier}. UTBoost, a test augmentation framework, finds deficient test cases on coding agent benchmarks that result in incorrect labels of task success and failure \citep{yu2025utboost,wang2025solved}. SWE-ABS strengthens tests with coverage and mutation tests and reveals false successes by evaluating detection of deliberately injected bugs \citep{yu2026sweabs}. Agents may also exploit visible tests rather than implement the intended behavior \citep{zhao2026specbench,zhong2025impossiblebench}. \benchmarkname therefore withholds its tests during the solve and injects them only after agent implementation. Our mutation analysis complements prior work by measuring which seeded faults are caught only by tests that run the full engine session.

\begin{table}[t]
\caption{Comparison of related coding and game-development benchmarks. $^\dagger$The generated program exposes the checked state.}
\label{tab:benchmark-comparison}
\centering
\small
\setlength{\tabcolsep}{4pt}
\begin{tabular}{@{}lcccc@{}}
\toprule
\textbf{Benchmark} & \textbf{Test-based} & \textbf{Runs the} & \textbf{Asserts on} & \textbf{Networked} \\
 & \textbf{behavioral verdict} & \textbf{game loop} & \textbf{runtime-owned state} & \textbf{client world}\\
\midrule
SWE-bench \citep{jimenez2023swebench} & \cmark & \xmark & \xmark & \xmark \\
AppForge \citep{ran2025appforge} & \cmark & \cmark & \xmark & \xmark \\
GameDevBench \citep{chi2026gamedevbench} & \cmark & \xmark & \cmark & \xmark \\
GameCraft-Bench \citep{gamecraft2026} & \xmark & \cmark & \xmark & \xmark \\
WorldCoder-Bench \citep{worldcoder2026} & \cmark & \cmark & \xmark\textsuperscript{$\dagger$} & \xmark \\
JAMER \citep{jamer2026} & \xmark & \cmark & \xmark & \xmark \\
PlayCoder \citep{peng2026playcoder} & \xmark & \cmark & \xmark & \xmark \\
AutoUE \citep{yin2026autoue} & \xmark & \cmark & \xmark & \xmark \\
\textbf{\benchmarkname} & \cmark & \cmark & \cmark & \cmark \\
\bottomrule
\end{tabular}
\end{table}

\section{\benchmarkname}

Each task in \benchmarkname is derived from a feature of a working UE5 project. Each task inherits the project's build and assets, and its reference solution is derived from the project's original implementation. This section describes how tasks are formulated, how the benchmark is constructed and calibrated, and what the evaluated task set covers. Section~\ref{sec:verification} describes the verification and evaluation pipeline in detail.

\subsection{Task Formulation}

The coding agent must implement the missing behavior according to the behavioral specification provided. The agent begins with a buildable project and the specification, and must integrate with the project's existing architecture to complete the task. After the agent finishes, its output is evaluated with the task's hidden suite of unit, integration, and end-to-end (E2E) tests (Section~\ref{sec:verification}). These tasks are substantially larger than those in issue-resolution benchmarks, with reference solutions adding a median of \geMedianSolutionLines{} lines across a median of \geMedianSolutionFiles{} files.

\begin{figure}[H]
\centering
\includegraphics[width=\linewidth]{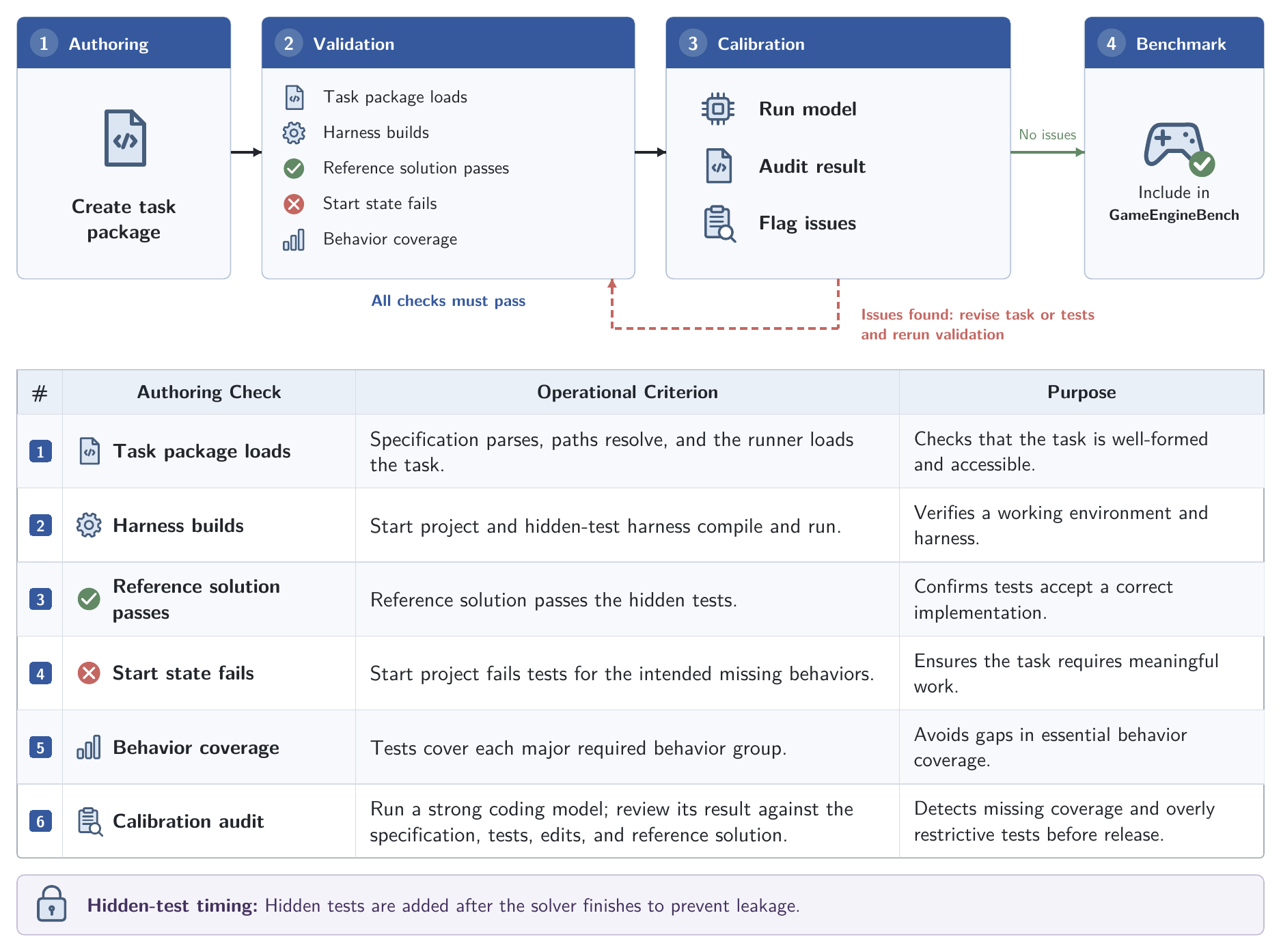}
\caption{Task authoring, test validation, LLM-as-a-Judge review, and task calibration process used while the tasks were built (Section~\ref{sec:scoring} describes the frozen suites used for scoring). Editable-file paths are an authoring aid and are not enforced by the grader.}
\label{fig:task-authoring}
\end{figure}

\subsection{Benchmark Construction}
\label{sec:construction}

Tasks are sourced from public UE5 repositories chosen for gameplay diversity and code quality. The tasks come from \numrepos{} projects: HordeTemplateV2Native, ActionRoguelike, Bomber, TargetVector, GASShooter, Eternal Crusade: Resurrection, UE-LASAA, PBCharacterMovement, and NanoGaussianSplatting. For each repository, we identify existing features in the codebase and remove their implementations from the source files. We then construct a behavioral specification that describes the missing feature. To solve the task, the model must reason through the existing C++ code and gameplay architecture rather than solving isolated programming problems.

To ensure the quality of our tasks, we used a multi-step authoring process with two distinct checks. \emph{Test validation} checks that the task package loads, the start project builds, the reference solution passes the tests, and the starting state fails those tests on the intended missing behaviors. \emph{LLM-as-a-Judge review} runs Claude Opus~4.8 on the task and asks GPT-5.5 as the judge to review the saved result against the behavior specification, test outcomes, model edits, and reference solution. This review identifies tests that are under-specified or over-constrained relative to the requested behavior. When issues are found, we revise the task or test suite and repeat the validation process. We refer to this iterative workflow as \emph{task calibration} (Figure~\ref{fig:task-authoring}). After this process, two game development experts with industry Unreal Engine experience manually reviewed each task for validity.

\subsection{Dataset Composition}

The evaluated set consists of \numtasks{} tasks focusing on gameplay and systems programming within existing codebases, covering gameplay mechanics, multiplayer behavior, AI and world orchestration, animation and character systems, UI and player-session systems, subsystem and plugin integration, runtime object management, data/configuration logic, movement systems, virtual- and augmented-reality (XR) behavior, and rendering-oriented plugin code. The benchmark is not designed to incorporate every facet of C++ game development. Coverage is intentionally lighter for domains such as audio, performance, platform support, security, and tooling.

\begin{figure}[t]
\centering
\includegraphics[width=.6\linewidth]{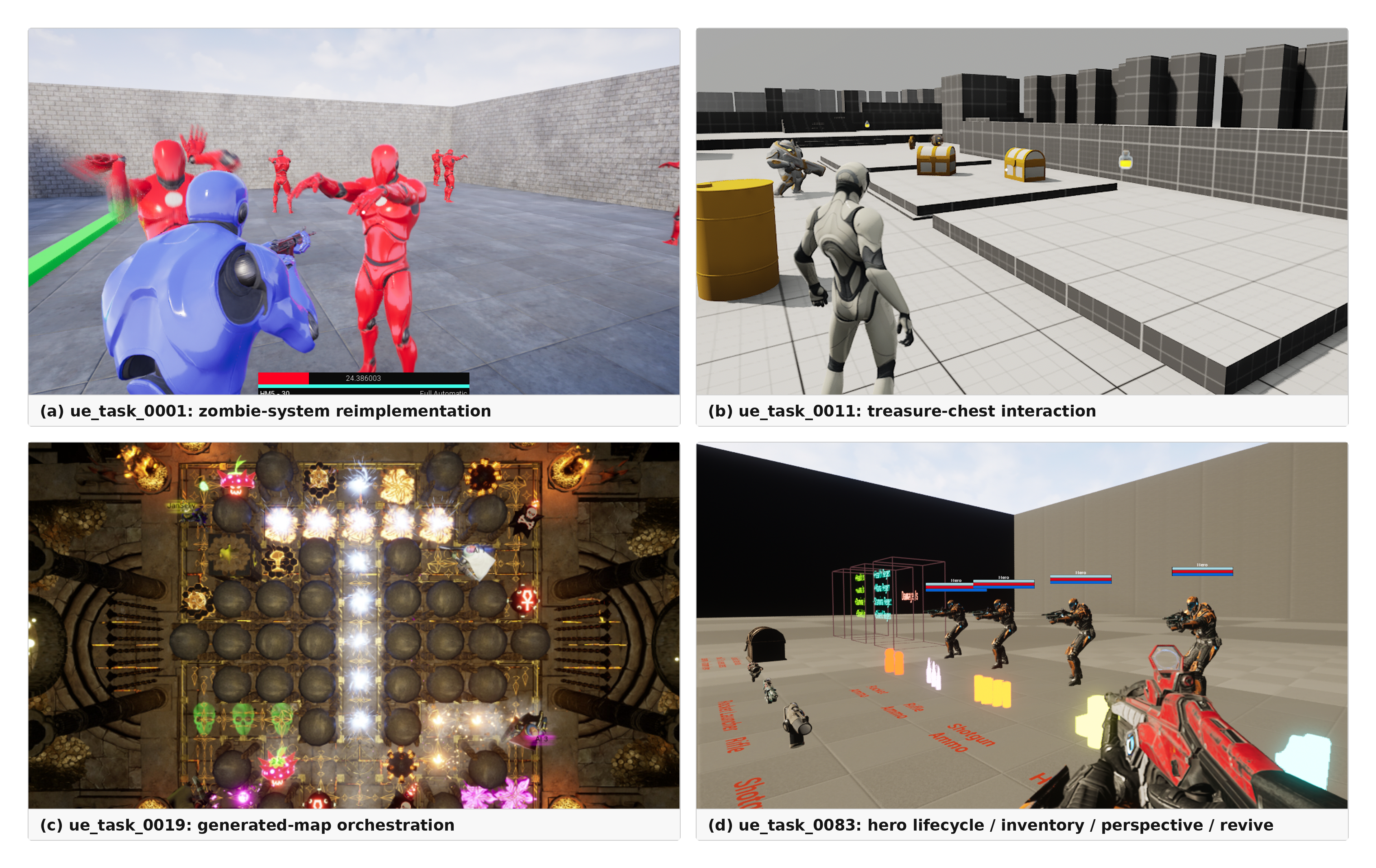}
\caption{Example tasks from GameEngineBench. Screenshots are taken from reference solutions to illustrate the gameplay and engine-system behaviors evaluated by the benchmark.}
\label{fig:task-gallery}
\end{figure}
  
Two sample tasks are described in Appendix~\ref{app:sample-tasks}.

\section{Verification and Evaluation}
\label{sec:verification}

Each task has a single hidden test suite (Section~\ref{sec:construction}), and the \numtasks{} test suites contain \geTagNumTests{} tests in total (three of them, in one task, are disabled and never run). Each test belongs to one of three tiers of increasing execution scope: \emph{unit} (\geNumTestsUnit{} tests), \emph{integration} (\geNumTestsIntegration{} tests in a synthetic world), and end-to-end (\emph{E2E}; \geNumTestsEtoE{} full-session tests); \geNumTestsStatic{} further test is purely static. Each test's tier, and whether it asserts on a client world, was labelled by two independent methods (Cohen's $\kappa=\geTagKappaTier$ for tier and $\geTagKappaWorld$ for world); the \geTagDisagreements{} disagreements were resolved by manual inspection. Static checks, which examine source declarations and metadata, are not separately scored. Rather, they are embedded in behavioral tests to affect verdicts. The design rule is to observe each requirement at the lowest tier that can fully observe it. Tiers therefore differ in observability of non-local properties, not strictness alone.

\subsection{Verification Tiers}

\emph{Static checks} inspect source declarations, annotations (such as \texttt{UPROPERTY} replication specifiers), and metadata without execution. They can verify that declarations exist, but cannot establish that behavior occurs. A solution may contain the expected function declaration while failing to initialize, update, or replicate the corresponding state. In the frozen suites, such assertions are embedded in unit, integration, and E2E tests and therefore affect the verdict: \geStaticOnlyRejections{} of the \geNumCells{} graded submissions fail only as a result of these static assertions.

\emph{Unit tests} execute functions or isolated object instances without a full game world, verifying local computations and constructor defaults. Without a running \texttt{UWorld}, however, they cannot observe temporal or distributed properties.

\emph{Integration tests} spawn actors in a synthetic game world, allowing Unreal Engine to manage component registration, subsystem initialization, and other single-world lifecycle behavior. This tier detects temporal violations, such as a component registered at the wrong lifecycle stage (e.g., in \texttt{BeginPlay}, invoked when an actor activates) or an actor querying a subsystem before initialization. However, it cannot observe distributed properties within a single world context.

\emph{E2E tests} execute a complete PIE session including the game loop, actors, and subsystems. In the \geNumTasksWithClients{} tasks that configure clients, this tier also verifies behavior across networked clients interacting through an authoritative server. The tests verify that client interactions propagate correctly and server state stays consistent. E2E tests observe both non-local-property types: lifecycle ordering and distributed consistency. In \geNumClientWorldTasks{} tasks, \geNumClientWorldTests{} tests directly inspect the client world to verify distributed behavior.

The distinction between tiers can be illustrated by a health feature. A static check can identify a replication annotation, a unit test can validate the damage calculation, and an integration test can verify that damage is applied to an actor in a world. Only a networked E2E test, however, can establish that the server produces the correct value, Unreal transmits it through its networking system, and a separate client receives the corresponding state. Our tiered design overcomes the limitations of evaluating specified runtime properties solely on source structure or isolated local behavior.

\subsection{Full-Session Verification}
\label{sec:full-session}

\textsc{Full-Session Verification} evaluates agent-generated code inside a complete PIE session in listen-server mode, with networked client worlds in \geNumTasksWithClients{} tasks. When configured in listen-server mode within one process, PIE creates an authoritative server world and separate client worlds connected via Unreal's networking system. Each world maintains its own actors, controllers, and state, and the verifier can inspect corresponding objects during the same test while respecting the ordinary property-replication and remote-procedure-call (RPC) mechanisms.

This arrangement provides a stronger oracle than inspecting source code, observing a single game instance, or asking the program to report its own state. \textsc{Full-Session Verification} can distinguish code that merely marks a specified property for replication from code that updates the property on the server, transmits it to the appropriate clients, and triggers the required client-side behavior. The same environment exposes lifecycle-dependent behavior involving initialization, possession, ticking (per-frame updates), callbacks, destruction, and cleanup. Therefore, the pipeline benefits not only from Unreal Engine's production-relevant native C++ environment, but also from PIE's ability to execute the complete gameplay lifecycle and directly observe state across networked worlds within an integrated testing environment. The verification pipeline consists of four stages.

\textbf{Test injection.} Solves and grading both run under the Harbor evaluation framework in Windows Docker containers. Each submission is graded in a fresh container started from the task's verifier image (the task's start state and the Unreal toolchain), separate from the container the agent worked in; the hidden tests are copied in only at this point. After the agent completes its implementation, a scan of the recorded trajectories of \geNumSolvesWithTrajectory{} of the \geNumCells{} graded solves found no access to test or solution paths; the other \geNumSolvesNoTrajectory{} solves ended at the agent time limit without a saved trajectory (Appendix~\ref{app:solver-env}). The grader injects the hidden test files into the project's test directory, replacing any test directory the agent left. The runner then adds the required test-module dependencies and compiles the submitted implementation alongside the hidden tests (Appendix~\ref{app:pipeline} gives the details of every stage). 

\textbf{Networked execution.} We use Unreal's automation framework to launch the project headlessly, except for \geNumRenderingTasks{} tasks requiring rendered output. The session runs in listen-server mode with $N$ additional client worlds in the same process, each with player controller and state, connected to the server over loopback (command line in Appendix~\ref{app:topology}). A total of \geNumTasksWithClients{} tasks configure at least one client ($N \geq 1$), and the other \geNumTasksNoClients{} run the server world only. Where applicable, tests are designed to control the starting conditions (e.g., fixed spawn locations and scripted actions), which reduces the dependence on manual input and random placement (Appendix~\ref{app:stability}).

\textbf{Distributed consistency.} In the \geNumClientWorldTasks{} tasks with client-world tests, we perform a specified action and wait for the relevant state to replicate to connected clients. The tests check whether the expected behavior occurs at each frame within an explicit time limit. Once detected, the tests then assert on the final states, order of events, and agreement between server and client worlds. Time-bounded evaluations are far more robust than relying on exact frame counts, which are susceptible to variability in frame durations and network delivery (Appendices~\ref{app:pipeline} and~\ref{app:stability}).

\textbf{Post-execution.} Finally, the runner records compilation status, per-test outcomes and assertion messages, automation logs, and the submitted workspace for audit.

\subsection{Evaluation Protocol and Scoring}
\label{sec:scoring}

Agent performance is measured by \emph{pass@1} (fraction of tasks solved). A submission's \emph{verdict} is based on whether the project successfully builds with the hidden tests injected and every test of the task's frozen suite passes. Crashes, missing results, and timeouts are counted as failures. Verdicts only come from these compiled tests (\emph{deterministic checks}), which removes the need for any model or human judgement. Regrading all \geNumCells{} submissions reproduced every verdict.

Attempts lost to infrastructure faults or provider usage limits were re-run under a fixed policy; every other attempt, including timeouts, was graded as it ended. Before the suites were frozen, tests and instructions were revised after inspecting failures from preliminary runs of the same \numconfigs{} configurations. After the suites were frozen, every configuration re-solved all \numtasks{} tasks, and only these new submissions are graded, so no graded submission was used for calibration.

Because the verdict relies exclusively on deterministic checks rather than model-based adjudication, the test suite is central to the validity of the benchmark. The test suite's design follows five key properties that enable reliable quantitative assessment of model outputs. First, tests are anchored to behavioral requirements rather than to the reference solution. Each task specification describes the behavioral requirements evaluated by the hidden tests, and \geNumTestsMapped{} of the \geTagNumTests{} tests are mapped to the requirements they observe. Hence, a test suite is designed to assert the behavior described by the requirement instead of the particular way the original solution was produced. Second, every reference solution passes its frozen suite (\geRefOfficialPass{} executions over two grading runs). Third, we measure what the suites detect by grading deliberately broken copies of the reference solutions (mutants), which are known to be wrong (Section~\ref{sec:verification-strength}). Fourth, rejections are attributable: in \geTestFailAttributable{} of the \geTestFailCells{} submissions that build, run, and fail at least one test, a failing test is mapped to a named requirement. Lastly, two experts independently reviewed 195 graded submissions (39 tasks drawn at random with a fixed seed, stratified by project and client-world type, each with five fixed configurations) against the task specification and behavioral contract. Out of the 195 submissions, Expert A reviewed 112 while Expert B reviewed 113, with 30 submissions independently reviewed by both. The experts initially agreed on 27 out of the 30 overlapping verdicts (90\%). The three disagreements arose from ambiguity in the task specification and uncertainty in interpreting the behavior. All three were resolved through discussion between the experts, which resulted in unanimous final verdicts. The independent review upheld every verdict.

\begin{table}[t]
\caption{Execution setup. Parentheses show reasoning effort (model computation budget).}
\label{tab:execution-setup}
\centering
\small
\setlength{\parindent}{0pt}
\setlength{\parskip}{0.6em}
\setlength{\tabcolsep}{5pt}
\begin{tabular}{@{}p{0.30\textwidth}p{0.60\textwidth}@{}}
\toprule
\textbf{Execution detail} & \textbf{Setup} \\
\midrule
Configurations (model + harness) & Claude Fable 5 (max), Claude Opus 4.8 (max, high), Claude Opus 4.7 (high), Claude Sonnet 4.6 (high) + Claude Code; GPT-5.5 (xhigh, high, medium) + Codex CLI; Gemini 3.1 Pro (High) + Antigravity CLI; Qwen 3.7 Plus and DeepSeek V4 Pro + Qwen Code
\\
Solver environment & Own Windows container per solve; prebuilt, read-only Unreal Engine 5.7.4 (an Installed Build) with its build toolchain; shell and file-editing tools; web access blocked; no version-control history; hidden tests and reference solution absent \\
Task mode & Unreal automation in PIE listen-server mode executed in a separate verifier container \\
Max solver time & \geTimeLimitSec{}\,s per solve\\
Verifier timeout & 3{,}600\,s per grade (build and tests)\\
Scoring & pass@1; one graded solve per task and configuration \\
\bottomrule
\end{tabular}

\end{table}

\section{Results}
\label{sec:results}
\begin{figure}[b]
\centering
\includegraphics[width=0.96\linewidth]{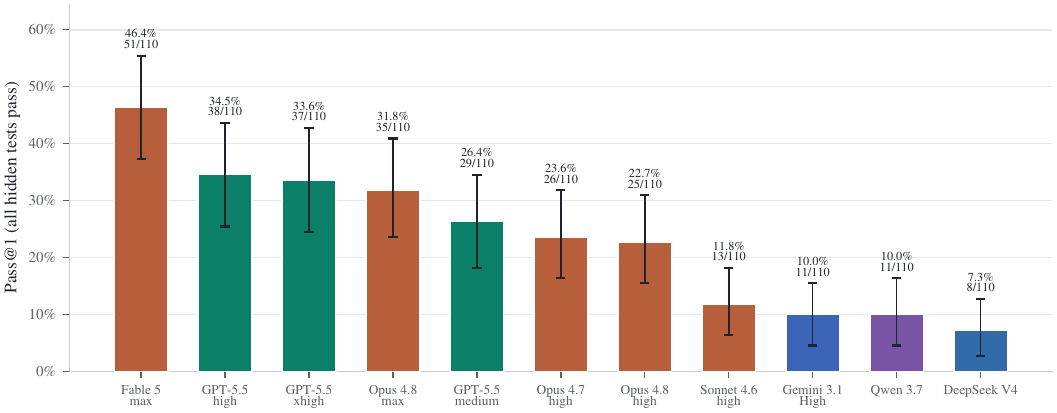}
\caption{pass@1 of each configuration on the \numtasks{} tasks, with 95\% confidence intervals (percentile bootstrap: tasks resampled with replacement, 10,000 resamples). A task counts as solved only if the submission compiles and passes every hidden test.}
\label{fig:main-success-rate}
\end{figure}

The results show that the benchmark is not saturated by current coding agents and remains difficult. The best-scoring configuration solves \bestpass{} of the tasks, and \numunsolved{} tasks are not solved by any configuration.
Figure~\ref{fig:main-success-rate} gives the main comparison. Claude Fable 5 (\texttt{max}) is numerically first at \gePassFableMax{} pass@1 (95\% confidence interval \gePassCIFableMax{}), followed by GPT-5.5 (\texttt{high}) at \gePassGptHigh{}, GPT-5.5 (\texttt{xhigh}) at \gePassGptXhigh{}, and Claude Opus 4.8 (\texttt{max}) at \gePassOpusFourEightMax{}. Paired tests (exact McNemar tests on the tasks that exactly one of two configurations solves, Holm-corrected for 55 comparisons) do not separate these four, so their order should be read as a tie. \geNumPairsDiffer{} of the \geNumPairs{} pairs of configurations differ significantly (Appendix~\ref{app:mcnemar}).

\subsection{Efficiency Tradeoffs}
Table~\ref{tab:cost} and Figure~\ref{fig:efficiency-frontier-main} (Appendix~\ref{app:supp-figures}) compare pass@1 with recorded cost and solve time. Claude Fable 5 (\texttt{max}) has the highest pass@1 and the highest recorded cost, \$\geCostPerPassFableMax{} per solved task, against \$\geCostPerPassGptHigh{} for GPT-5.5 (\texttt{high}), with which it is statistically tied. The configurations with the highest capability are therefore not the most cost efficient. The Gemini harness records neither cost nor tokens, so Gemini is not compared on cost.

\subsection{Solved Sets Are Complementary}
Figure~\ref{fig:coverage-complementarity} (Appendix~\ref{app:supp-figures}) shows which configurations solve which tasks. Solved sets are mostly, but not fully, nested. Together the \numconfigs{} configurations solve \geNumSolvedByAny{} of the \numtasks{} tasks, against \gePassCountFableMax{} for the best single configuration, and \geNumSingleSolver{} tasks are solved by exactly one configuration (examples in Appendix~\ref{app:complementarity-examples}). Because each configuration attempts each task once, part of this pattern may reflect solve-to-solve variance.

\subsection{Failure Mechanisms}\label{sec:failure-mechanisms}

\begin{figure}[t]
\centering
\includegraphics[width=\linewidth]{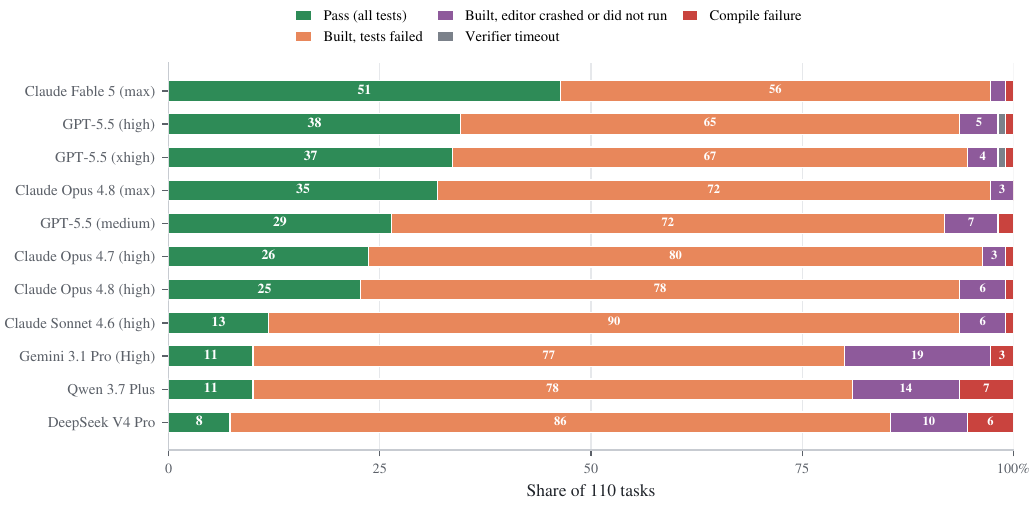}
\caption{Outcome decomposition per configuration over its \numtasks{} graded submissions: pass; built but failed at least one hidden test; built, but the editor running the hidden tests crashed or did not run; verifier timeout; compile failure.}
\label{fig:failure-compile}
\end{figure}

Figure~\ref{fig:failure-compile} breaks down how submissions fail. Compile failures are rare, as shown in only \geNumCompileFail{} of the \geNumCells{} submissions (\geCompileFailClaudeCode{} under Claude Code, \geCompileFailCodex{} under Codex, \geCompileFailAntigravity{} under Antigravity, \geCompileFailQwenCode{} under Qwen Code). Most failing submissions (\geNumBuiltTestFail{}) build and run but fail at least one hidden test. Task~\texttt{19} (Appendix~\ref{app:sample-tasks}) exemplifies this common failure mode: every configuration builds a submission, and all \numconfigs{} fail its only test, an E2E test; in ten of them, the server and the client do not end up with the same layout. The code compiles and resembles plausible Unreal C++, but its behavior across worlds is inconsistent. We also classify failures by the property each failing test checks, using the mutant classes (tests were classified with LLM assistance). Of the \geNumBuiltTestFail{} submissions that build and fail at least one test, \gePropFailNonLocal{} (\gePropFailNonLocalPct\%) fail a test of a non-local property. \gePropFailOrdering{} failed an ordering test, \gePropFailDistributed{} failed a distributed test, and \gePropFailCrossFile{} failed a cross-file test, while \gePropFailLocalLogic{} fail a local-logic test. Submissions can fail multiple non-local properties at once; 180 fail tests of two or more. Non-local properties are therefore involved in most real test failures.

\subsection{Verification Strength}\label{sec:verification-strength}

We characterize what the frozen suites detect with seeded faults (\emph{mutants}), which are copies of the reference solution with a single fault injected at one site. Mutants were drafted with LLM assistance and reviewed, and only those reviewed as nonequivalent and valid are used. Each of the \geMutApplicable{} mutants from \geMutNumTasks{} tasks is graded using its corresponding task's full frozen test suite, producing a killed/not-killed verdict for each mutant. We classify mutants by the property the fault violates: \emph{local logic} (an error inside one function), \emph{cross-file} (an inconsistency between files or classes), \emph{ordering} (work done at the wrong point of the lifecycle), \emph{distributed} (a replication or server-authority error), and \emph{hollow} (declarations kept, behavior removed). Because these mutants were also used to find and fix weak tests, the kill rate is optimistic (an upper bound) for new, unseen bugs. 

\begin{table*}[t]
\centering
\small
\caption{Mutation analysis on the frozen suites: each of the \geMutApplicable{} valid mutants (\geMutNumTasks{} tasks) is graded with its task's full frozen suite, and Full suite is the share of mutants it rejects (kills). Unit, Integ., and E2E give the share of mutants with at least one failing test of that tier within the same suite; ``missed by U+I'' counts mutants killed by an E2E test but by no unit or integration test (``S+U+I'' also excludes the single standalone static test), and ``only by client world'' counts those whose failing tests all assert on a client world. ``Other'' pools \geMutOtherN{} mutants from 12 small classes, and all rates are in-sample upper bounds (Section~\ref{sec:verification-strength}).}
\label{tab:mutation}
\label{tab:mutation-sensitivity}%
\resizebox{\textwidth}{!}{%
\begin{tabular}{lrrrrrrrrr}
\toprule
 & & \multicolumn{4}{c}{Kill rate (\%)} & \multicolumn{2}{c}{Killed, but missed by} & Caught only & \\
\cmidrule(lr){3-6}\cmidrule(lr){7-8}
Mutant class & $n$ & Full suite & Unit & Integ. & E2E & U+I & S+U+I & by client world & Survivors \\
\midrule
Local logic & 207 & 97.1 & 32.9 & 50.7 & 42.0 & 45 & 45 & 9 & 6 \\
Cross-file & 91 & 96.7 & 15.4 & 45.1 & 53.8 & 35 & 35 & 8 & 3 \\
Ordering & 76 & 94.7 & 19.7 & 42.1 & 51.3 & 28 & 28 & 10 & 4 \\
Distributed & 67 & 97.0 & 0.0 & 11.9 & 85.1 & 55 & 54 & 51 & 2 \\
Hollow & 110 & 98.2 & 28.2 & 49.1 & 41.8 & 28 & 28 & 8 & 2 \\
Other (12 classes) & 16 & 100.0 & 18.8 & 62.5 & 18.8 & 3 & 3 & 1 & 0 \\
\midrule
All & 567 & 97.0 & 23.1 & 44.1 & 49.6 & 194 & 193 & 87 & 17 \\
\bottomrule
\end{tabular}
}
\end{table*}

\textbf{The full test suite comprehensively catches the vast majority of mutants.} The full suite rejects \geMutKilled{} of the \geMutApplicable{} mutants (\geMutKillRate{}; Table~\ref{tab:mutation}). Of the \geMutDistN{} distributed mutants, \geMutDistKilled{} are killed; \geMutDistMissedByUI{} of these are caught by no unit or integration test, and \geMutDistClientOnly{} of those \geMutDistMissedByUI{} only by assertions on a client world. The table also demonstrates that deeper tiers do not catch more faults in general. For local-logic mutants, integration tests successfully fail more often (\geMutLocalKillRateIntegration{}) than E2E tests (\geMutLocalKillRateEtoE{}). E2E tests are not supersets of lower tiers: each requirement is tested at the lowest tier that can observe it, so local logic is mostly checked by unit and integration tests. The \geMutSurvivors{} surviving mutants are characterized in Appendix~\ref{app:mutation-survivors}, where we conclude that most survive because an assertion or a fixture input cannot tell them from the reference.

\textbf{Shallower tests accept failing submissions.}\label{sec:real-submissions} Mutants do not necessarily resemble the errors agents make, so we measure the deeper tests on the \geNumCells{} graded submissions (Appendix~\ref{app:real-submissions}). \geEtoEOnlyFails{} of the \geEtoEOnlyWouldBe{} submissions that pass every static, unit, and integration test fail at least one E2E test (\geEtoEOnlyPct{}; \geEtoEOnlyIntFails{} of \geEtoEOnlyIntWouldBe{} on tasks whose suites also contain integration tests). Of the \geSWPassTotal{} submissions that pass every test confined to a single world, \geSWPassCWFail{} (\geSWPassCWFailPct{}) fail an assertion on a client world. Static declaration and reflection assertions, which are embedded in behavioral tests, decide few verdicts on their own (\geStaticOnlyRejections{} of \geNumCells{}).

\textbf{Rankings based on shallow verifiers mask failure.} Without E2E tests, each configuration's pass@1 would be inflated by \geEtoEOnlyDeltaMin--\geEtoEOnlyDeltaMax{} percentage points and the rankings would change (Table~\ref{tab:e2e-only}). Without client-world tests alone, pass@1 would be inflated by \geCWDeltaMin--\geCWDeltaMax{} points and the ranking of \geNumRankChangesWithoutCW{} configurations would change (Table~\ref{tab:qkua-q1}).

\section{Limitations and Next Steps}
While the current benchmark provides a strong foundation for evaluating model capabilities across a broad range of gameplay-facing tasks, including combat, inventory, networking, actor lifecycle, player/session flows, subsystem integration, and data/configuration behavior, the task coverage is incomplete. Audio, memory, performance, editor tools, build systems, and security-sensitive systems remain underrepresented. This distribution partly reflects the availability of public Unreal projects, which more commonly expose gameplay code than engine modifications, rendering work, platform integrations, performance infrastructure, or security-sensitive systems. Future versions should expand these areas while preserving the benchmark's emphasis on evaluating engine-managed behavior. \benchmarkname also evaluates model--harness configurations, and harness differences (tools, shells, output limits) are not separated from model effects. Agents were instructed not to run the editor or automation tests, so the results measure implementation under compile-time feedback only. Each configuration solves each task once, so solve-to-solve variance is not measured. 

\section{Conclusion}
\benchmarkname marks a concrete step toward advancing coding-agent evaluation from source-level plausibility to verified behavior within a running, stateful system. The \numtasks{}-task set requires agents to integrate missing functionality into existing Unreal Engine projects, while hidden tests evaluate the resulting implementations with tests that range from unit checks to full networked sessions. At the highest tier, \textsc{Full-Session Verification} executes the complete gameplay lifecycle in PIE and directly inspects engine-managed state. In \geNumClientWorldTasks{} tasks, the pipeline is used to verify behavior in a separate client world, exposing failures that compilation, source inspection, and single-world tests cannot observe. The results show that current model--harness configurations frequently produce plausible and compilable code without reproducing the required runtime behavior. These findings demonstrate the importance of evaluating generated code within the lifecycle, subsystem, and networking contexts in which it must operate. Although the measured results are specific to Unreal Engine and the evaluated task set, the verification principle applies more broadly: code for stateful frameworks should be tested within the runtime contexts that define its behavior. \benchmarkname provides both a challenging benchmark for native game-engine development and a foundation for evaluating coding agents in other lifecycle-dependent and distributed software systems.

\subsection*{AI use statement}

Large language models (LLMs) were used in this work. Claude Opus~4.8 helped implement the grader and write the hidden tests, and LLM agents triaged failures and drafted the test and instruction revisions made before the suites were frozen. During task authoring, GPT-5.5 served as the calibration judge (Section~\ref{sec:construction}). Mutants were also drafted with LLM assistance, and one of the two test-tagging methods is an LLM classifier. LLMs assisted with analysis scripts, figures (Figure~\ref{fig:task-authoring} was drawn with an image-generation model), and writing. No LLM contributes to any reported verdict. Experts reviewed 195 graded submissions (Section~\ref{sec:scoring}) and upheld every verdict. The authors take responsibility for the final content of this work.

\subsection*{Ethics statement}

The benchmark involves no human subjects or personal data. Tasks are derived from public Unreal Engine projects whose licences vary (Appendix~\ref{app:licensing}). Upstream material is not redistributed, and setup scripts fetch each project from its public repository and apply our task changes.

\subsection*{Reproducibility statement}

Section~\ref{sec:verification} and Appendix~\ref{app:verification-methods} describe the tests, the grading pipeline, and verdict stability, and Appendix~\ref{app:licensing} describes the release (grader, frozen suites, task list, per-submission verdicts, mutants, and setup scripts that rebuild each task from its upstream source). Verdicts are reproducible given a submission, but agent solves are not deterministic.

\bibliographystyle{iclr2027_conference}
\bibliography{references}

\newpage
\appendix
\raggedbottom
\section{Sample Tasks}
\label{app:sample-tasks}
We show two sample tasks that no evaluated configuration solves to provide a better idea of what a task entails. Each example demonstrates a failure mode that coding agents struggle with.

Task~\texttt{6} (Round-Based Match Loop) is a multiplayer task in HordeTemplate, a cooperative horde-survival game for one to four players on a dedicated server (a server process without a local player). The agent must implement the repeating match cycle: a pause phase during which the trader accepts purchases, an active round that ends when its timer expires or all enemies are dead, the end of the match after the last round followed by the server moving all players to the next map in the rotation, an immediate level reset when all players die, and the refusal of a match end while enemies remain. The round number, pause state, timers, and remaining-enemy count must replicate to clients, and state changes must happen only on the server. The reference solution changes five source files, mostly the game state. \geExMatchLoopNearMiss{} configurations pass all but one of the \geExMatchLoopTests{} tests, but no configuration passes the E2E test that plays a two-round match in a networked session; in \geExMatchLoopAuthorityFails{} submissions, that test observes a client starting the pause phase, which only the server may do.

Task \texttt{19} (Generated Map Orchestrator) is narrower in file count (the reference changes two source files) but still integration-heavy. When a round starts, the grid must be filled with the correct level actors in their cells, and two clients in the same session must see identical maps. The cell assignment must not block the game thread (the thread that runs gameplay), regeneration must reuse pooled actors instead of spawning new ones, exactly one ready signal must fire per regeneration on each peer (the server and each client) after all of that peer's actors are present, and map-size changes at runtime must replicate to all clients. Similarly, this task is difficult because the agent must coordinate generation, actor pooling, and timing while handling proper syncing. All \numconfigs{} configurations build a submission, but none passes the task's single E2E test (Section~\ref{sec:failure-mechanisms}).

\section{Test Construction and Full-Session Execution}
\label{app:verification-methods}

This section describes how the benchmark derives tasks, constructs behavioral oracles, executes full Play-in-Editor (PIE) sessions, and controls sources of variation.

\subsection{Task Derivation and Workspace}

Each task is derived from a feature in a working Unreal Engine 5 project. Task authors identify an existing behavior, derive the reference solution from its original implementation, and remove the implementation required to produce that behavior while retaining a buildable starting project. The public task specification describes externally observable requirements rather than prescribing a particular implementation.

Agents may edit any file. The hidden tests and verifier-owned fixtures are added only at grading time, replacing any test directory the agent left (Appendix~\ref{app:pipeline}).

Hidden tests and reference solutions are stored outside the solver container, and the tests are injected only after the agent completes its implementation, in a separate verifier container. The agent implementation and injected tests are then compiled together using the project's Unreal build configuration. Compilation is a measured stage because it exposes errors involving C++, headers, reflection macros, generated code, module boundaries, and engine API dependencies.

\subsection{Requirement-to-Oracle Construction}

The design rule is that task authors decompose each behavioral specification into observable requirements, construct at least one oracle for each, and assign each oracle to the lowest-fidelity verifier that can observe the property, so that a source-level proxy does not stand in for runtime behavior.

\begin{table}[t]
\centering
\small
\setlength{\tabcolsep}{4pt}
\begin{tabularx}{\textwidth}{@{}lXXX@{}}
\toprule
\textbf{Tier} & \textbf{Execution context} & \textbf{Representative observations} & \textbf{Properties not established} \\
\midrule
Static & Source and class metadata inspection & Declarations, signatures, macro annotations, and constructor-defined metadata & Executed behavior, lifecycle, and networking \\
\addlinespace
Unit & Direct calls on functions or isolated objects & Pure logic, defaults, conversions, and local state transitions & World lifecycle and distributed state \\
\addlinespace
Integration & Actors and components in a synthetic \texttt{UWorld} & Registration, initialization, subsystem interaction, and single-world lifecycle & Cross-client delivery and authority consistency \\
\addlinespace
End-to-end & Complete PIE session & Tick-dependent behavior, possession, authority, RPCs, replication, and client-visible effects & Deployment properties not represented by PIE \\
\bottomrule
\end{tabularx}
\caption{Verification tiers and the properties observable at each level of execution fidelity. In the frozen suites, static checks are assertions inside unit, integration, and E2E tests (\geNumTestsStatic{} test is purely static).}
\label{tab:verification-tier-details}
\end{table}

Static checks can establish that a required declaration or annotation exists, but they cannot establish that the associated behavior executes correctly. Unit tests evaluate local computations without a complete game world. Integration tests create engine-managed objects in a synthetic \texttt{UWorld} (Unreal's game-world object), exposing initialization and lifecycle behavior within one context. End-to-end tests execute a complete PIE session and are required when correctness depends on the game loop, authority, remote procedure calls, replication, or state in a separate client world. A task may use several tiers when its specification combines local, lifecycle-dependent, and distributed requirements.

\subsection{Networked PIE Topology}
\label{app:topology}

PIE starts a live game session from the project loaded by the Unreal Editor \citep{epic2026piemultiplayer}. For networked evaluation, the grader configures a listen server and the required client worlds and enables execution under one process. The server world is authoritative and also contains the listen server's local player. Each client has its own \texttt{UWorld}, actors, \texttt{APlayerController}, \texttt{AGameState}, and \texttt{APlayerState} (engine-managed world, controller, game state, and player state objects).

Worlds communicate over loopback via Unreal's networking system (UNetDriver, UNetConnection, and replicated properties), including remote procedure calls. The single-process topology simplifies observation but does not replace replication: state reaches a client only if the implementation produces it on the authority (the server) and Unreal transmits it to that client. Because all worlds are accessible within one process, a test running on the game thread can locate corresponding server and client objects and compare their state without log scraping or a custom inter-process protocol.

The grader uses an invocation of the following form, with task-specific values substituted for the project, test filter, and client count:

\par\medskip
\begingroup
\setlength{\fboxsep}{9pt}
\setlength{\fboxrule}{0.4pt}
\noindent\fcolorbox{black!20}{black!3}{%
\begin{minipage}{\dimexpr\linewidth-2\fboxsep-2\fboxrule\relax}
{\ttfamily\small
UnrealEditor <project>.uproject\\
\hspace*{2em}-ExecCmds="Automation RunTests <filter>; Quit"\\
\hspace*{2em}-Unattended -CrashForUAT -NullRHI -NoSplash -nocef\\
\hspace*{2em}-NumClients=N -ListenServer\par}
\end{minipage}%
}
\endgroup
\par\medskip

The flag \texttt{-NullRHI} disables rendering, and \texttt{-NumClients} and \texttt{-ListenServer} set up the networked session; the command line also includes \texttt{-nocef}. Tasks that need rendered output replace \texttt{-NullRHI} with an off-screen rendering configuration, and task-specific values are recorded with each task configuration. PIE exercises the engine's gameplay lifecycle and networking, but it is not identical to a packaged build, a dedicated-server deployment, or communication over a wide-area network.

\subsection{Full-Session Verification Pipeline}
\label{app:pipeline}

The verification pipeline consists of four stages. Section~\ref{sec:full-session} summarizes them; this section is the detailed description.

\noindent\textbf{Test injection and compilation.}
After the agent completes its implementation, grading runs in a separate verifier container: the hidden tests replace any test directory the agent left, verifier-owned fixtures are applied, and \texttt{Build.cs} is updated with the required test-module dependencies. The project's Editor target is then compiled with the injected tests (unity builds disabled), and a compilation failure is recorded as an evaluation outcome rather than bypassed.

\noindent\textbf{Session initialization.}
The runner launches the project through Unreal's automation framework. For networked tasks, it starts PIE in listen-server mode with the required client count and same-process execution enabled. The test waits until the required worlds, controllers, actors, and subsystems are available instead of assuming that they exist on the first frame.

\noindent\textbf{Stimulus, convergence, and assertion.}
The test establishes controlled initial conditions and performs the action specified by the task, such as applying damage on the authority, beginning a round, possessing an actor, invoking an RPC, or destroying an object. A latent automation command (repeats until ready) yields control to the engine between polls. The command polls a readiness condition until the required state appears or a time budget expires. Once the state has converged, the test evaluates task-specific invariants in the relevant worlds.

In \geNumClientWorldTasks{} tasks (\geNumClientWorldTests{} tests), assertions directly inspect a separate client world. A server-side value therefore does not pass merely because it was computed correctly on the authority; the test also requires the corresponding client state or client-side effect to appear. A schematic replication oracle is:

\par\medskip
\begingroup
\setlength{\fboxsep}{9pt}
\setlength{\fboxrule}{0.4pt}
\noindent\fcolorbox{black!20}{black!3}{%
\begin{minipage}{\dimexpr\linewidth-2\fboxsep-2\fboxrule\relax}
{\ttfamily\small
start networked PIE\\
wait until server and client actors exist\\
apply the action to the server actor\\
while (state unconverged) and time remains:\\
\hspace*{2em}yield to next engine tick\\
assert client state equals expected state\par}
\end{minipage}%
}
\endgroup
\par\medskip

\noindent\textbf{Result collection and cleanup.}
After execution, the runner parses Unreal's automation log to recover per-test verdicts, using the editor's JSON test report for results the log lost on exit. A submission fails if the build fails, the editor does not run or crashes, the number of completed tests differs from the number discovered, no test is found, any test fails, or the grade exceeds the verifier timeout. The complete log, test results, and submitted workspace are retained for auditing.

\subsection{Controls for Verdict Stability}
\label{app:stability}

Unreal Engine is not a fixed-step deterministic simulator. Frame durations, AI updates, physics, navigation, and loopback delivery may complete on different ticks across repeated executions. We therefore target verdict stability rather than identical execution traces.

Tests are designed to use controlled initial conditions (fixed positions, scripted actions, and explicit gameplay-state transitions) and to poll state predicates with explicit time budgets instead of fixed sleeps; \geNumTimingSensitiveTests{} tests are nevertheless sensitive to host load. Assertions emphasize final-state invariants, event ordering, monotonic progress, and cross-world agreement rather than exact frame counts.

The grader runs every grade in a fresh verifier container with a standardized command line produced by one script shared by all tasks; task-specific changes are recorded in the task configuration.

\texttt{-NullRHI} disables frame rasterization but preserves the game loop, actor lifecycle, gameplay logic, and networking. It is therefore appropriate for behavioral tests that do not require rendered output. The \geNumRenderingTasks{} tasks whose tests require rendered output declare a rendering-enabled configuration that removes \texttt{-NullRHI} and renders off-screen (Appendix~\ref{app:verification-methods}, command line).

\textbf{Measured stability.} Regrading every submission on fresh machines reproduced all \geNumCells{} verdicts (\gePerTestFlips{} of \gePerTestCompared{} individual test results differed, none changing a verdict), and repeat grades of 100 submissions changed none (Table~\ref{tab:stability-compact}). Every reference solution passed in both grading runs (\geRefOfficialPass{}); the only failed executions were two first attempts of one load-sensitive test, which passed when re-run on an idle machine.

\begin{table}[t]
\centering
\small
\caption{Verdict stability: a flip is a verdict that differs between two gradings of the same submission. Rows compare a full regrade on fresh machines with the verdicts first recorded (split by whether the first grading used the same or an earlier grader), 100 repeat grades with the same grader, and two independent gradings of each reference solution; ``confirmed'' counts are after re-running failed executions on an idle machine.}
\label{tab:stability-compact}
\begin{tabular}{lrr}
\toprule
Set & Raw flips & Confirmed flips \\
\midrule
Identical-condition cells & 0/121 & 0/121 \\
Older-grader cells & 0/1{,}089 & 0/1{,}089 \\
All submission cells & 0/1{,}210 & 0/1{,}210 \\
Repeat grades (same grader) & 0/100 & 0/100 \\
Reference r1 vs r2 & 0/110 & -- \\
Reference executions passing & 220/222 & 220/220 \\
\bottomrule
\end{tabular}
\end{table}

\subsection{Task Validation and Calibration}

During authoring, each task was subjected to positive and negative controls (Figure~\ref{fig:task-authoring}). The task package must load, the starting project must compile, the reference solution must pass every required test, and the incomplete starting state must fail on the behaviors intentionally removed. A failure caused only by an invalid fixture, environment configuration, or unrelated project defect does not validate the oracle. Authoring-time failures were inspected through execution logs and test messages

\textbf{Calibration of the reported suites.} Before the official graded run, tests and instructions were revised after inspecting preliminary run failures, with LLM agents helping to triage failures and draft fixes. The revisions fixed grader and environment problems, corrected tests that the reference failed or that over-constrained valid implementations, edited one reference solution, and clarified instructions; after the suites were frozen, all configurations re-solved all tasks.

We additionally evaluate verifier sensitivity using deliberately incorrect implementations, or mutants, each graded with the full frozen suite (Section~\ref{sec:verification-strength}). This analysis characterizes verifier depth; it does not establish that the test suite captures every possible incorrect implementation.

\subsection{Recorded Outputs and Scope}

For each evaluated run, the runner records compilation status, per-test outcomes and assertion messages, automation logs, and the submitted workspace. A run passes only when compilation succeeds and every test of the frozen suite passes (Section~\ref{sec:scoring}). Agent trajectories are not released.

Full-Session Verification establishes behavior within the tested PIE configuration. It does not by itself establish rendering correctness under \texttt{-NullRHI}, performance at production scale, equivalence to packaged builds or dedicated-server fleets, resilience to packet loss or wide-area latency, platform-specific behavior, or correctness for interaction paths absent from the test suite. These limitations bound the experimental claim: the method evaluates whether an implementation satisfies specified invariants during a controlled and auditable engine session.

\noindent\begin{minipage}{\linewidth}
\subsection{Solver Environment}
\label{app:solver-env}

Table~\ref{tab:solver-env} summarizes the solver environment, from the executed command lines and the recorded agent trajectories. No agent ran automation tests or a PIE session, and none of the \geNumWebAttempts{} network attempts, made in 140 solves, succeeded. A writable shared cache of derived shader data was also used by grading runs and may have held shaders compiled from reference solutions; the one solve that tried to read such data failed and is graded as a failure.

\begin{table}[H]
\centering
\small
\renewcommand{\arraystretch}{1.15}
\caption{Solver environment of the graded solves (110 per configuration): each agent ran in its own Windows container with a shell, file tools, and a read-only Unreal Engine 5.7.4 build, without hidden tests, reference solution, version history, or web access, and was told it could compile but not run the editor or tests. The columns count solves that built, that built only a non-Editor target, that launched the editor or ran tests, that made network attempts (successful ones in parentheses), that obtained version history, and that timed out, followed by compile failures and passes in the reported grading.}
\label{tab:solver-env}
\begin{tabularx}{\linewidth}{@{}lXll@{}}
\toprule
Configuration & Harness & Effort & Web tools \\
\midrule
Claude Fable 5 (max) & Claude Code 2.1.283 & max & disallowed \\
Claude Opus 4.8 (max) & Claude Code 2.1.283 & max & disallowed \\
Claude Opus 4.8 (high) & Claude Code 2.1.283 & high & disallowed \\
Claude Opus 4.7 (high) & Claude Code 2.1.283 & high & disallowed \\
Claude Sonnet 4.6 (high) & Claude Code 2.1.283 & high & disallowed \\
GPT-5.5 (xhigh) & Codex CLI 0.158.0 & xhigh & disabled \\
GPT-5.5 (high) & Codex CLI 0.158.0 & high & disabled \\
GPT-5.5 (medium) & Codex CLI 0.158.0 & medium & disabled \\
Gemini 3.1 Pro (High) & Antigravity CLI 1.2.12 & (High) variant & deny hook \\
Qwen 3.7 Plus & Qwen Code 0.24.6 & thinking & excluded \\
DeepSeek V4 Pro & Qwen Code 0.24.6 & thinking & excluded \\
\bottomrule
\end{tabularx}
\par\medskip
\footnotesize
\setlength{\tabcolsep}{2pt}
\begin{tabularx}{\linewidth}{@{}l*{8}{>{\centering\arraybackslash}X}@{}}
\toprule
Configuration & Built & \shortstack{Non-\\Editor} & \shortstack{Editor/\\tests} & \shortstack{Solves\\with net.\\att.} & \shortstack{Git\\hist.} & Timeout & \shortstack{Compile\\fail} & Passes \\
\midrule
Claude Fable 5 (max) & 110 & 0 & 1/0 & 54 (0) & 0 & 2 & 1 & 51 \\
Claude Opus 4.8 (max) & 110 & 1 & 0/0 & 1 (0) & 0 & 1 & 0 & 35 \\
Claude Opus 4.8 (high) & 107 & 0 & 0/0 & 6 (0) & 0 & 0 & 1 & 25 \\
Claude Opus 4.7 (high) & 107 & 1 & 0/0 & 0 (0) & 0 & 0 & 1 & 26 \\
Claude Sonnet 4.6 (high) & 103 & 21 & 0/0 & 0 (0) & 0 & 1 & 1 & 13 \\
GPT-5.5 (xhigh) & 110 & 6 & 0/0 & 27 (0) & 0 & 0 & 1 & 37 \\
GPT-5.5 (high) & 110 & 12 & 0/0 & 17 (0) & 0 & 0 & 1 & 38 \\
GPT-5.5 (medium) & 110 & 14 & 0/0 & 6 (0) & 0 & 0 & 2 & 29 \\
Gemini 3.1 Pro (High) & 104 & 9 & 0/0 & 8 (0) & 0 & 6 & 3 & 11 \\
Qwen 3.7 Plus & 106 & 47 & 0/0 & 0 (0) & 0 & 2 & 7 & 11 \\
DeepSeek V4 Pro & 106 & 17 & 0/0 & 21 (0) & 0 & 0 & 6 & 8 \\
\bottomrule
\end{tabularx}
\end{table}

\end{minipage}

\clearpage
\section{Supplemental Result Figures}
\label{app:supp-figures}
Below are extra figures for analysis, and the cost table (Table~\ref{tab:cost}). All figures use the reported verdicts of the same \numtasks{} tasks and \numconfigs{} configurations.

\noindent\begin{minipage}{\linewidth}
\begin{table}[H]
\centering
\fontsize{7.5}{8.5}\selectfont
\setlength{\tabcolsep}{2pt}
\caption{Cost and efficiency of the graded solves, one per configuration and task; costs are API list-price equivalents, and cost per pass is total cost divided by passes. $^\dagger$Computed from token counts and list prices rather than recorded by the harness (for DeepSeek V4 Pro, an assumed cache-hit price of 10\% of the input price dominates the estimate). n.r.: not reported by the Antigravity harness.}
\label{tab:cost}
\begin{tabular*}{\linewidth}{@{\extracolsep{\fill}}lrrrrrrl@{}}
\toprule
Configuration & \shortstack{pass@1\\(\%)} & \shortstack{Median\\time\\(min)} & \shortstack{Mean\\cost\\(\$)} & \shortstack{Median\\cost\\(\$)} & \shortstack{Cost per\\pass\\(\$)} & \shortstack{Mean\\output tok.\\(k)} & Cost source \\
\midrule
Claude Fable 5 (max) & 46.4 & 36.8 & 19.56 & 18.32 & 42.20 & 116.3 & recorded by harness \\
GPT-5.5 (high) & 34.5 & 15.8 & 2.42 & 2.29 & 7.01 & 17.4 & recorded by harness \\
GPT-5.5 (xhigh) & 33.6 & 21.6 & 4.02 & 3.98 & 11.94 & 29.4 & recorded by harness \\
Claude Opus 4.8 (max) & 31.8 & 41.9 & 10.72 & 10.21 & 33.70 & 146.4 & recorded by harness \\
GPT-5.5 (medium) & 26.4 & 14.6 & 1.90 & 1.80 & 7.22 & 12.2 & recorded by harness \\
Claude Opus 4.7 (high) & 23.6 & 19.3 & 5.50 & 4.77 & 23.27 & 42.0 & recorded by harness \\
Claude Opus 4.8 (high) & 22.7 & 23.4 & 5.08 & 4.57 & 22.36 & 62.0 & recorded by harness \\
Claude Sonnet 4.6 (high) & 11.8 & 24.5 & 3.11 & 2.72 & 26.28 & 72.5 & recorded by harness \\
Gemini 3.1 Pro (High) & 10.0 & 15.2 & n.r. & n.r. & n.r. & n.r. & not reported \\
Qwen 3.7 Plus & 10.0 & 23.7 & 0.33$^\dagger$ & 0.32$^\dagger$ & 3.23$^\dagger$ & 47.2 & computed from tokens$^\dagger$ (108/110) \\
DeepSeek V4 Pro & 7.3 & 21.9 & 1.19$^\dagger$ & 1.16$^\dagger$ & 16.40$^\dagger$ & 39.2 & computed from tokens$^\dagger$ \\
\bottomrule
\end{tabular*}
\end{table}

\begin{figure}[H]
\centering
\includegraphics[width=0.95\linewidth]{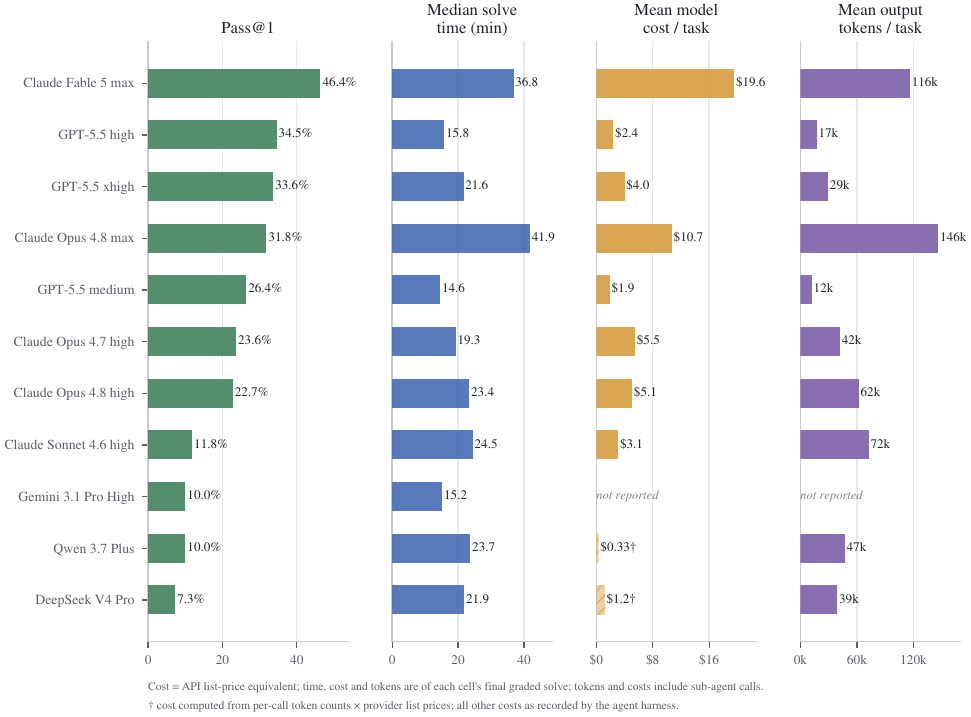}
\caption{pass@1, median solve time, mean cost per task, and mean output tokens per task for each configuration. Costs are API list-price equivalents recorded by the agent harness; costs marked $\dagger$ are computed from token counts and list prices (Table~\ref{tab:cost}); the Gemini harness reports neither cost nor tokens.}
\label{fig:efficiency-frontier-main}
\end{figure}
\end{minipage}

\begin{figure}[H]
\centering
\begin{subfigure}[t]{0.44\linewidth}
  \centering
  \begin{tikzpicture}
  \begin{axis}[
    width=\linewidth,height=6cm,
    ybar,bar width=7pt,bar shift=0pt,
    xmin=-0.7,xmax=11.7,ymin=0,ymax=48,
    xtick={0,1,...,11},
    ytick={0,10,20,30,40},
    xlabel={Number of configurations\\that solve a task},ylabel={Number of tasks},
    xlabel style={align=center},
    axis lines=left,axis line style={black!45},
    tick style={black!45},tick label style={font=\scriptsize},
    label style={font=\footnotesize},
    ymajorgrids=true,grid style={black!10},
    nodes near coords,point meta=y,
    every node near coord/.append style={font=\scriptsize,anchor=south},
    clip=false,
  ]
    \addplot[fill=black!45,draw=none] coordinates {(0,43)};
    \addplot[fill={rgb,255:red,84;green,121;blue,91},draw=none] coordinates {(1,16) (2,9) (3,5) (4,10) (5,9) (6,1) (7,4) (8,6) (9,4) (10,1) (11,2)};
  \end{axis}
  \end{tikzpicture}
  \caption{Tasks grouped by how many of the 11 configurations solve them.}
\end{subfigure}\hfill
\begin{subfigure}[t]{0.54\linewidth}
  \centering\includegraphics[width=\linewidth,height=6cm,keepaspectratio]{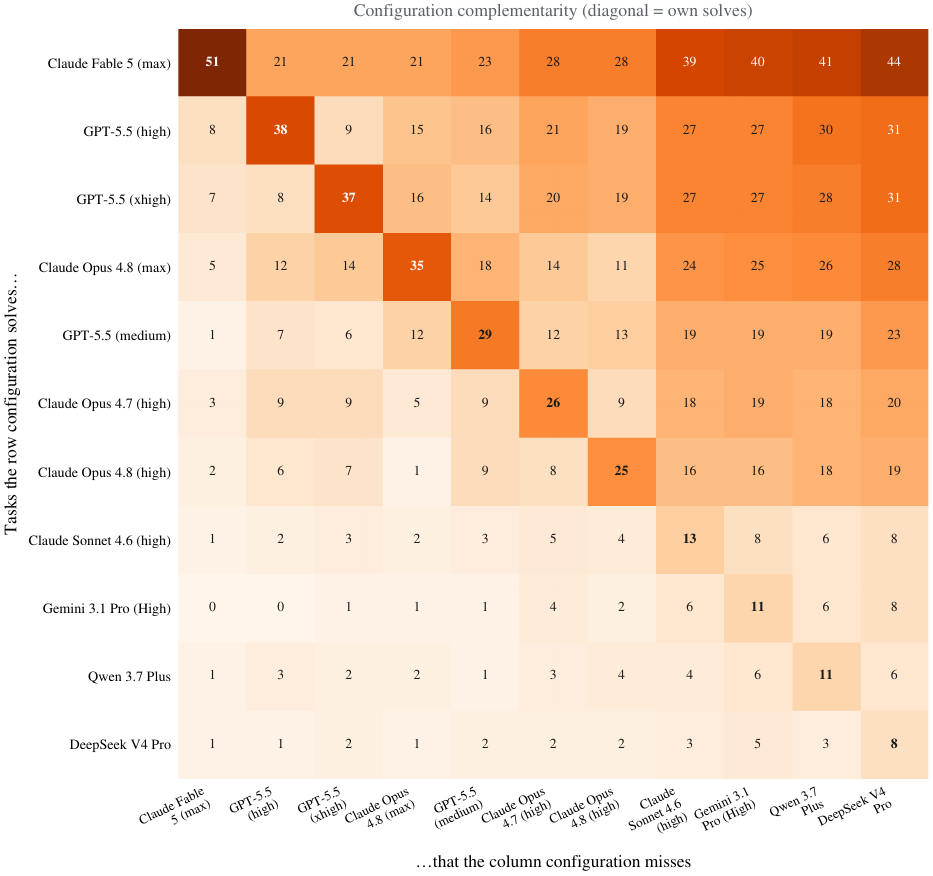}
  \caption{Pairwise task complementarity: tasks the row configuration solves that the column configuration misses (diagonal: own solves)}.
\end{subfigure}\\[0.5em]
\begin{subfigure}[t]{0.49\linewidth}
  \centering\includegraphics[width=\linewidth,height=3.8cm,keepaspectratio]{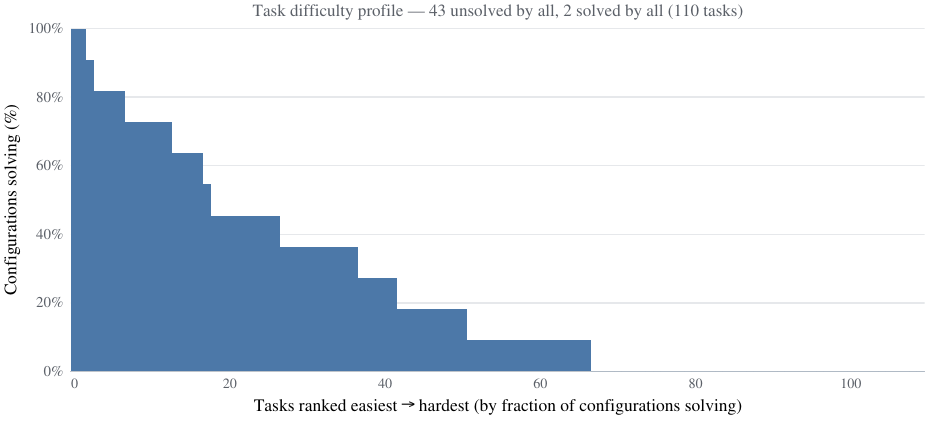}
  \caption{Tasks ranked by fraction of configurations that solve them.}
\end{subfigure}\hfill
\begin{subfigure}[t]{0.49\linewidth}
  \centering\includegraphics[width=\linewidth,height=3.8cm,keepaspectratio]{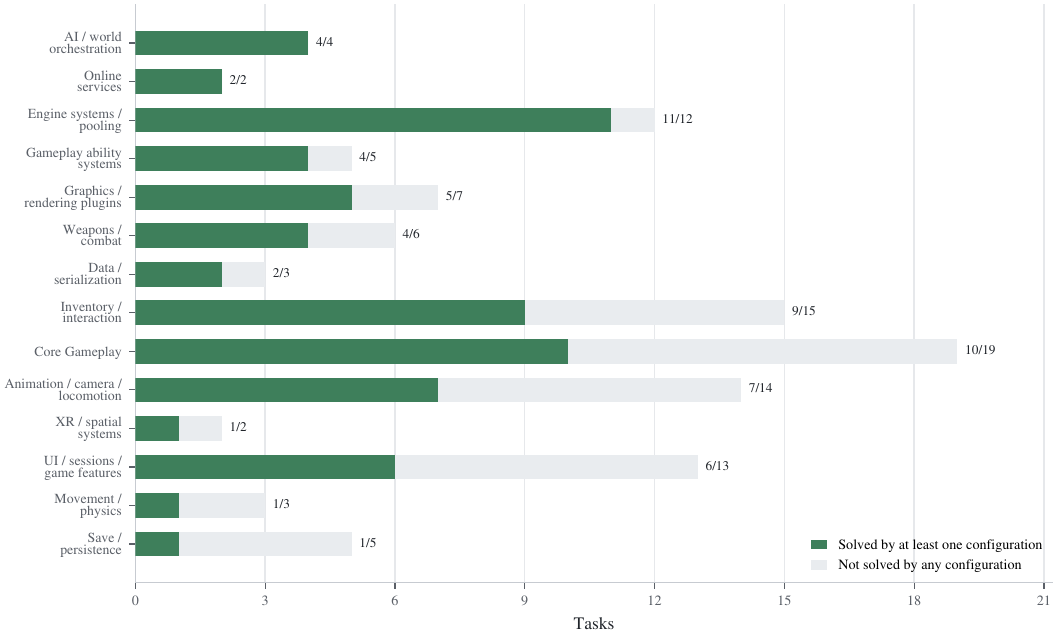}
  \caption{Tasks solved by at least one configuration (green) vs.\ none (gray), by area.}
\end{subfigure}\\[0.5em]
\begin{subfigure}[t]{\linewidth}
  \centering\includegraphics[width=\linewidth]{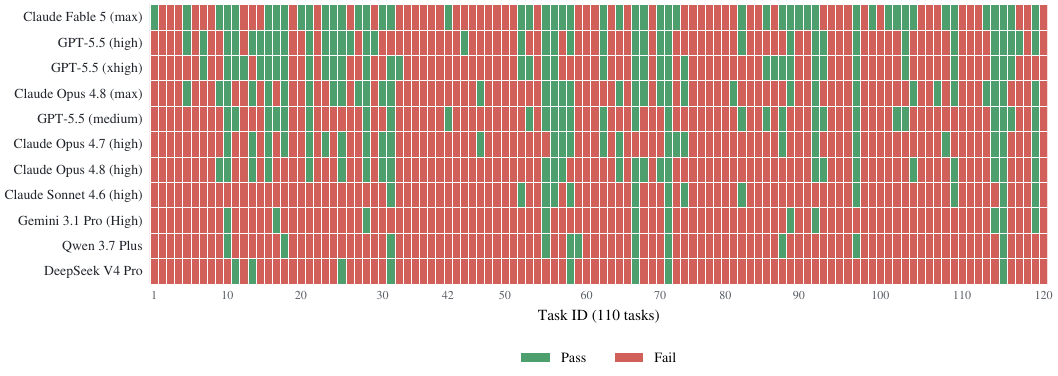}
  \caption{Per-task outcome matrix over the \numtasks{} tasks for all \numconfigs{} configurations; green is pass.}
\end{subfigure}
\caption{Task coverage and model complementarity. All panels use the verdicts of the deterministic checks (compiles and passes every hidden test) for the same \numtasks{} tasks and \numconfigs{} configurations, taken from one set of verdict records. Panel (e) is the task-level view behind the aggregate pass@1 ranking.}
\label{fig:coverage-complementarity}
\end{figure}

\begin{figure}[H]
\centering
\begin{subfigure}[t]{0.49\linewidth}
  \centering\includegraphics[width=\linewidth]{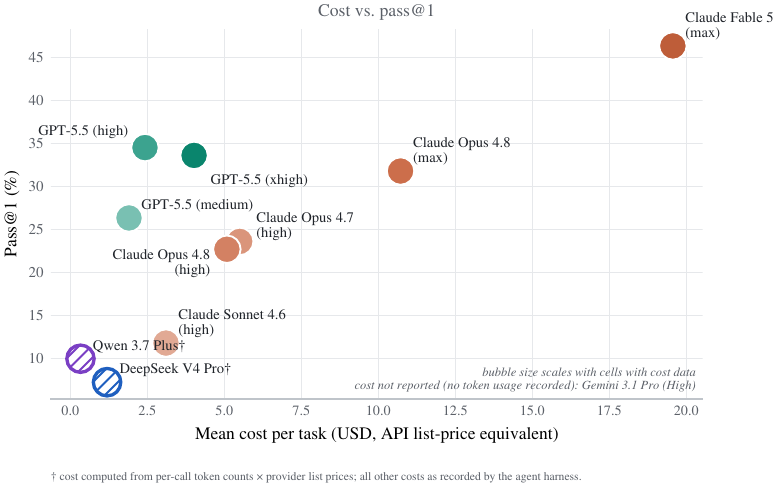}
  \caption{Cost vs.\ pass rate, where cost data is available (hatched: computed cost, $\dagger$)}.
\end{subfigure}\hfill
\begin{subfigure}[t]{0.49\linewidth}
  \centering\includegraphics[width=\linewidth]{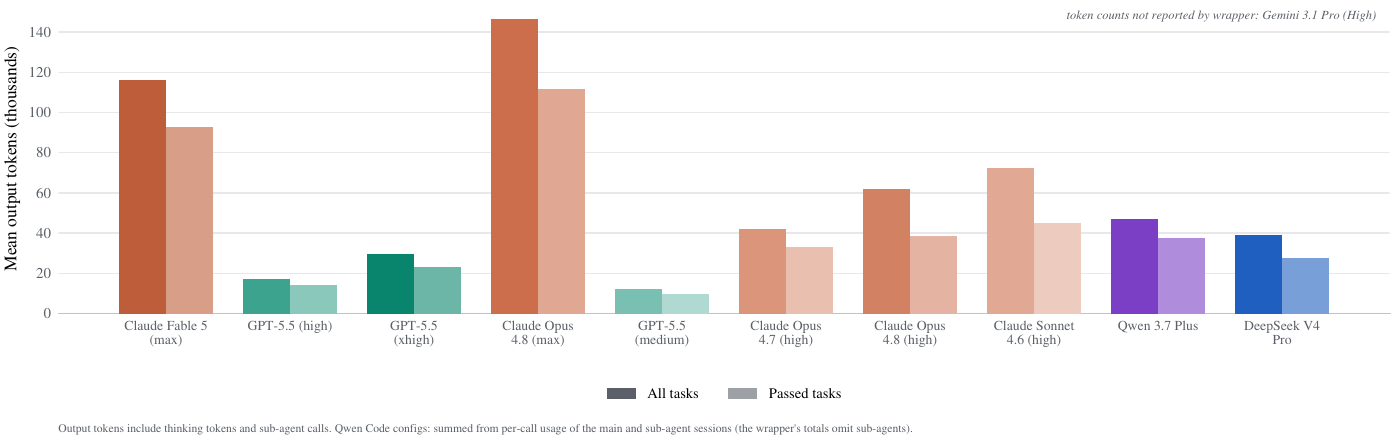}
  \caption{Output tokens per task, over all tasks and over passed tasks}.
\end{subfigure}\\[0.5em]
\begin{subfigure}[t]{0.49\linewidth}
  \centering\includegraphics[width=\linewidth]{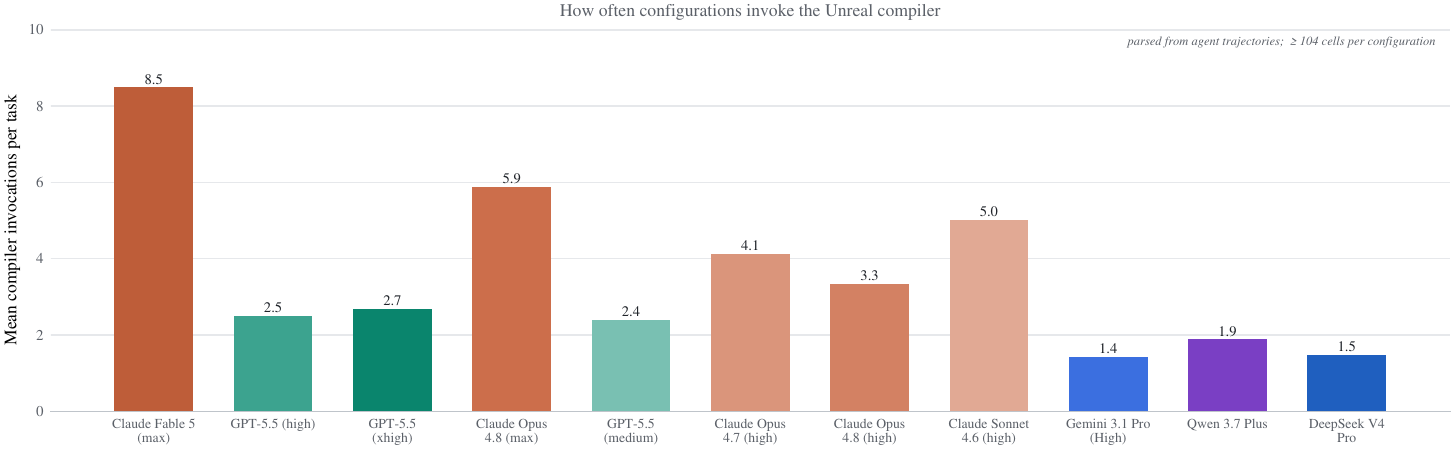}
  \caption{Mean compiler invocations per task.}
\end{subfigure}\hfill
\begin{subfigure}[t]{0.49\linewidth}
  \centering\includegraphics[width=\linewidth]{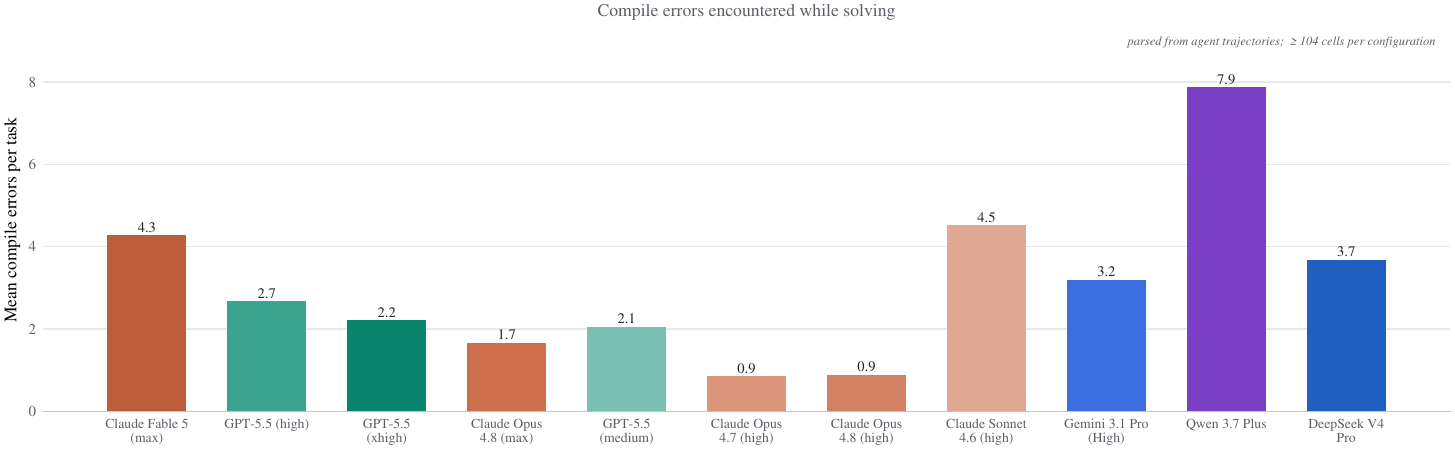}
  \caption{Mean compile errors during solving.}
\end{subfigure}
\caption{Efficiency and resource-use diagnostics. Compile counts in (c) and (d) are parsed from the agent trajectories (at least 104 solves per configuration), so they are only as complete as each harness's trajectory. Costs in (a) are recorded by the harness except where marked $\dagger$, and the Gemini harness reports neither cost nor tokens, so (a) and (b) omit it.}
\label{fig:app-diagnostics}
\end{figure}
\section{Client-World and E2E-Only Failures on Real Submissions}
\label{app:real-submissions}
Tables~\ref{tab:qkua-q1}, \ref{tab:client-world-split}, and~\ref{tab:e2e-only} give the per-configuration results behind Section~\ref{sec:real-submissions}.

\begin{table}[!htbp]
\centering
\small
\caption{Client-world assertions on real submissions: of the \geNumCells{} submissions, \geSWPassTotal{} pass every single-world test, and \geSWPassCWFail{} of these (\geSWPassCWFailPct{}) fail only on client-world assertions (CW-fail). ``w/o client-world'' is pass@1 with client-world assertions dropped, and $\Delta$ is the resulting change in rank.}
\label{tab:qkua-q1}
\begin{tabular}{lrrrrrr}
\toprule
 & \multicolumn{2}{c}{pass@1 (\%)} & SW-pass / & \multicolumn{3}{c}{Rank} \\
\cmidrule(lr){2-3}\cmidrule(lr){5-7}
Configuration & All tests & w/o client-world & CW-fail & All tests & w/o CW & $\Delta$ \\
\midrule
Claude Fable 5 (max) & 46.4 & 51.8 & 6 & 1 & 1 & 0 \\
GPT-5.5 (high) & 34.5 & 39.1 & 5 & 2 & 2= & 0 \\
GPT-5.5 (xhigh) & 33.6 & 38.2 & 5 & 3 & 4 & $-1$ \\
Claude Opus 4.8 (max) & 31.8 & 39.1 & 8 & 4 & 2= & $+2$ \\
GPT-5.5 (medium) & 26.4 & 31.8 & 6 & 5 & 5 & 0 \\
Claude Opus 4.7 (high) & 23.6 & 30.9 & 8 & 6 & 6 & 0 \\
Claude Opus 4.8 (high) & 22.7 & 30.0 & 8 & 7 & 7 & 0 \\
Claude Sonnet 4.6 (high) & 11.8 & 16.4 & 5 & 8 & 8 & 0 \\
Gemini 3.1 Pro (High) & 10.0 & 14.5 & 5 & 9= & 9 & 0 \\
Qwen 3.7 Plus & 10.0 & 13.6 & 4 & 9= & 10 & $-1$ \\
DeepSeek V4 Pro & 7.3 & 10.9 & 4 & 11 & 11 & 0 \\
\midrule
Total submissions & 284 & 348 & 64 & & & \\
\bottomrule
\end{tabular}
\end{table}

\begin{table}[!htbp]
\centering
\small
\caption{pass@1 (\%) on tasks with $\geq$1 client-world test vs.\ the rest.}
\label{tab:client-world-split}
\begin{tabular}{lrr}
\toprule
Configuration & Client-world tasks & Other tasks \\
 & ($n=53$) & ($n=57$) \\
\midrule
Claude Fable 5 (max) & 41.5 & 50.9 \\
GPT-5.5 (high) & 30.2 & 38.6 \\
GPT-5.5 (xhigh) & 32.1 & 35.1 \\
Claude Opus 4.8 (max) & 24.5 & 38.6 \\
GPT-5.5 (medium) & 26.4 & 26.3 \\
Claude Opus 4.7 (high) & 17.0 & 29.8 \\
Claude Opus 4.8 (high) & 18.9 & 26.3 \\
Claude Sonnet 4.6 (high) & 5.7 & 17.5 \\
Gemini 3.1 Pro (High) & 7.5 & 12.3 \\
Qwen 3.7 Plus & 3.8 & 15.8 \\
DeepSeek V4 Pro & 5.7 & 8.8 \\
\bottomrule
\end{tabular}
\end{table}

\begin{table}[H]\centering\small
\caption{Submissions rejected only by E2E tests: they build and pass every static, unit, and integration test but fail at least one E2E test. Overall, \geEtoEOnlyFails{} of the \geEtoEOnlyWouldBe{} would-be passes (\geEtoEOnlyPct{}) are such failures; on tasks whose suite also has integration tests, \geEtoEOnlyIntFails{} of \geEtoEOnlyIntWouldBe{} (\geEtoEOnlyIntPct{}).}
\label{tab:e2e-only}
\begin{tabular}{@{}lrrrrr@{}}\toprule
Configuration & pass@1 & E2E-only fails & pass@1 w/o E2E & $\Delta$ (pp) & rank $\rightarrow$ \\ \midrule
Claude Fable 5 (max) & 46.4 & 19 & 63.6 & +17.3 & 1$\rightarrow$1 \\
GPT-5.5 (high) & 34.5 & 18 & 50.9 & +16.4 & 2$\rightarrow$3 \\
GPT-5.5 (xhigh) & 33.6 & 19 & 50.9 & +17.3 & 3$\rightarrow$3 \\
Claude Opus 4.8 (max) & 31.8 & 23 & 52.7 & +20.9 & 4$\rightarrow$2 \\
GPT-5.5 (medium) & 26.4 & 17 & 41.8 & +15.5 & 5$\rightarrow$6 \\
Claude Opus 4.7 (high) & 23.6 & 19 & 40.9 & +17.3 & 6$\rightarrow$7 \\
Claude Opus 4.8 (high) & 22.7 & 22 & 42.7 & +20.0 & 7$\rightarrow$5 \\
Claude Sonnet 4.6 (high) & 11.8 & 20 & 30.0 & +18.2 & 8$\rightarrow$8 \\
Gemini 3.1 Pro (High) & 10.0 & 20 & 28.2 & +18.2 & 9$\rightarrow$9 \\
Qwen 3.7 Plus & 10.0 & 14 & 22.7 & +12.7 & 9$\rightarrow$11 \\
DeepSeek V4 Pro & 7.3 & 18 & 23.6 & +16.4 & 11$\rightarrow$10 \\
\midrule All (1{,}210 submissions) & 23.5 & 209 & 40.7 & +17.3 & \\
\bottomrule\end{tabular}\end{table}

\section{Complementarity Examples}
\label{app:complementarity-examples}
Figure~\ref{fig:coverage-complementarity} shows that capability is mostly, but not completely, nested: \geNumSingleSolver{} tasks are solved by exactly one configuration, from \geNumSingleSolverConfigs{} different configurations.

Two such tasks, whose tests were not revised during calibration, illustrate this. In Task~\texttt{26} (Pool Manager editor nodes), only GPT-5.5 (\texttt{high}) passes all tests. Of the other ten submissions, four crash or fail to compile before the tests run, and six fail at least one test; in Claude Fable 5 (\texttt{max}), for example, the array node's output cannot connect to a typed array, so Blueprints that use it do not compile.

In Task~\texttt{107} (replication-graph routing), only Claude Opus 4.8 (\texttt{max}) passes all tests; eight of the other ten configurations fail at least one of its two client-world tests of owner relevancy, in which the owning client must keep its pawn while the server points the player's view at a distant actor.

\clearpage
\section{Surviving Mutants}
\label{app:mutation-survivors}
Table~\ref{tab:mutation-survivors} lists the \geMutSurvivors{} mutants that the frozen suites do not reject (Section~\ref{sec:verification-strength}), with the reason each survives. For example, in one task a bot-spawn timer was changed from repeating to one-shot, so bots stop spawning after the first wave; the test checks only that the timer is active immediately after match start and never advances time past a second interval, so the mutant passes.

\begin{table}[H]
\centering
\small
\caption{Property coverage of the hidden tests. Each executed test (\geCovNumTests{} in total) is assigned the property class it checks; a task covers a class if at least one of its tests does. \geCovTasksAnyNonLocal{} of the \numtasks{} tasks test at least one non-local property and \geCovTasksTwoPlusNonLocal{} test two or more.}
\label{tab:property-coverage}
\begin{tabular}{lrr}
\toprule
Property class & Tests & Tasks covered \\
\midrule
Local logic & \geCovTestsLocal{} & \geCovTasksLocal{} \\
Cross-file & \geCovTestsCrossFile{} & \geCovTasksCrossFile{} \\
Ordering & \geCovTestsOrdering{} & \geCovTasksOrdering{} \\
Distributed & \geCovTestsDistributed{} & \geCovTasksDistributed{} \\
\midrule
\multicolumn{3}{l}{Tasks by number of non-local classes covered: none \geCovTasksNoNonLocal{}, one \geCovTasksOneNonLocal{}, two \geCovTasksTwoNonLocal{}, three \geCovTasksThreeNonLocal{}} \\
\bottomrule
\end{tabular}
\end{table}

\begin{table}[H]
\centering
\fontsize{7.5}{8.5}\selectfont
\caption{The \geMutSurvivors{} mutants that the frozen suite does not reject, with the reason it misses each, assigned by reading the patch, the instruction, and the tests. Weak assertion: the mutated code runs, but no assertion or fixture input distinguishes it; untested path: no test exercises it; fixture limit: the distinguishing condition cannot be produced deterministically.}
\label{tab:mutation-survivors}
\begin{tabular}{@{}llp{0.56\textwidth}l@{}}
\toprule
Task & Mutant & Class: injected fault & Why it survives \\
\midrule
0011 & m2 & Local logic: drive the lid with SetWorldRotation instead of SetRelativeRotation & weak assertion \\
0013 & m2 & Local logic: timer loop flag true changed to false & weak assertion \\
0013 & m4 & Hollow: asset load and spawn dispatch removed after budget deduction & weak assertion \\
0014 & m2 & Local logic: wrong receiver: snapshot the corpse mesh pose instead of the source mesh pose & weak assertion \\
0014 & m5 & Local logic: drop branch: ImpulseBoneRemapping lookup before AddImpulseAtLocation & weak assertion \\
0014 & m6 & Cross-file: bypass subsystem: World->SpawnActor instead of URogueActorPoolingSubsystem::AcquireFromPool & weak assertion \\
0022 & m5 & Cross-file: drop the InitBomb hookup in the spawn callback & weak assertion \\
0022 & m6 & Ordering: read the avatar's activation-time cell instead of the event-time cell & fixture limit \\
0037 & m2 & Hollow: remove pawn visuals update from ApplyChosenMeshData, keep the broadcast & weak assertion \\
0046 & m6 & Distributed: multicast feedback dropped: authority branch of StartImpactReaction calls StartImpactReactionImplementation locally instead of MulticastStartImpactReaction & untested path \\
0047 & m3 & Cross-file: compare readiness against the wrong gameplay tag & untested path \\
0060 & m8 & Ordering: removal notification moved after ownership teardown in NotifyPlayerDestroyed & weak assertion \\
0079 & m7 & Ordering: move the FirstStreamRequestSinceMapLoad blocking upgrade below the UGameplayStatics::LoadStreamLevel call in LoadStreamLevel & weak assertion \\
0083 & m7 & Local logic: wrong member: third-person camera FOV reset uses Default1PFOV & weak assertion \\
0110 & m2 & Local logic: minimum correspondence count reduced from three to two & weak assertion \\
0112 & m4 & Ordering: state update moved before validation: bIsCrouched cleared before the encroachment overlap test & untested path \\
0112 & m6 & Distributed: deleted network-prediction block: uncrouch MeshTranslationOffset adjustment removed & weak assertion \\
\bottomrule
\end{tabular}
\end{table}

\clearpage
\section{Pairwise Significance Tests}
\label{app:mcnemar}
Table~\ref{tab:mcnemar-pairs} lists all paired exact McNemar tests behind the comparisons in Section~\ref{sec:results}.

\begin{table}[H]
\centering
\fontsize{7.5}{8.5}\selectfont
\setlength{\tabcolsep}{4pt}
\caption{All \geNumPairs{} pairwise comparisons on the \numtasks{} tasks (paired exact McNemar tests); ``A only'' and ``B only'' count tasks solved by one configuration but not the other. Decisions use Holm-adjusted $p$ values at $\alpha=0.05$: \geNumPairsDiffer{} pairs differ and \geNumPairsTie{} are ties.}
\label{tab:mcnemar-pairs}
\begin{tabular}{llrrrrl}
\toprule
Config A & Config B & A only & B only & $p$ (exact) & $p$ (Holm) & Decision \\
\midrule
Claude Fable 5 (max) & GPT-5.5 (high) & 21 & 8 & 0.024 & 0.482 & tie \\
Claude Fable 5 (max) & GPT-5.5 (xhigh) & 21 & 7 & 0.013 & 0.275 & tie \\
Claude Fable 5 (max) & Claude Opus 4.8 (max) & 21 & 5 & 0.003 & 0.070 & tie \\
Claude Fable 5 (max) & GPT-5.5 (medium) & 23 & 1 & $<$0.001 & $<$0.001 & different \\
Claude Fable 5 (max) & Claude Opus 4.7 (high) & 28 & 3 & $<$0.001 & $<$0.001 & different \\
Claude Fable 5 (max) & Claude Opus 4.8 (high) & 28 & 2 & $<$0.001 & $<$0.001 & different \\
Claude Fable 5 (max) & Claude Sonnet 4.6 (high) & 39 & 1 & $<$0.001 & $<$0.001 & different \\
Claude Fable 5 (max) & Gemini 3.1 Pro (High) & 40 & 0 & $<$0.001 & $<$0.001 & different \\
Claude Fable 5 (max) & Qwen 3.7 Plus & 41 & 1 & $<$0.001 & $<$0.001 & different \\
Claude Fable 5 (max) & DeepSeek V4 Pro & 44 & 1 & $<$0.001 & $<$0.001 & different \\
GPT-5.5 (high) & GPT-5.5 (xhigh) & 9 & 8 & 1.000 & 1.000 & tie \\
GPT-5.5 (high) & Claude Opus 4.8 (max) & 15 & 12 & 0.701 & 1.000 & tie \\
GPT-5.5 (high) & GPT-5.5 (medium) & 16 & 7 & 0.093 & 1.000 & tie \\
GPT-5.5 (high) & Claude Opus 4.7 (high) & 21 & 9 & 0.043 & 0.770 & tie \\
GPT-5.5 (high) & Claude Opus 4.8 (high) & 19 & 6 & 0.015 & 0.307 & tie \\
GPT-5.5 (high) & Claude Sonnet 4.6 (high) & 27 & 2 & $<$0.001 & $<$0.001 & different \\
GPT-5.5 (high) & Gemini 3.1 Pro (High) & 27 & 0 & $<$0.001 & $<$0.001 & different \\
GPT-5.5 (high) & Qwen 3.7 Plus & 30 & 3 & $<$0.001 & $<$0.001 & different \\
GPT-5.5 (high) & DeepSeek V4 Pro & 31 & 1 & $<$0.001 & $<$0.001 & different \\
GPT-5.5 (xhigh) & Claude Opus 4.8 (max) & 16 & 14 & 0.856 & 1.000 & tie \\
GPT-5.5 (xhigh) & GPT-5.5 (medium) & 14 & 6 & 0.115 & 1.000 & tie \\
GPT-5.5 (xhigh) & Claude Opus 4.7 (high) & 20 & 9 & 0.061 & 1.000 & tie \\
GPT-5.5 (xhigh) & Claude Opus 4.8 (high) & 19 & 7 & 0.029 & 0.551 & tie \\
GPT-5.5 (xhigh) & Claude Sonnet 4.6 (high) & 27 & 3 & $<$0.001 & $<$0.001 & different \\
GPT-5.5 (xhigh) & Gemini 3.1 Pro (High) & 27 & 1 & $<$0.001 & $<$0.001 & different \\
GPT-5.5 (xhigh) & Qwen 3.7 Plus & 28 & 2 & $<$0.001 & $<$0.001 & different \\
GPT-5.5 (xhigh) & DeepSeek V4 Pro & 31 & 2 & $<$0.001 & $<$0.001 & different \\
Claude Opus 4.8 (max) & GPT-5.5 (medium) & 18 & 12 & 0.362 & 1.000 & tie \\
Claude Opus 4.8 (max) & Claude Opus 4.7 (high) & 14 & 5 & 0.064 & 1.000 & tie \\
Claude Opus 4.8 (max) & Claude Opus 4.8 (high) & 11 & 1 & 0.006 & 0.158 & tie \\
Claude Opus 4.8 (max) & Claude Sonnet 4.6 (high) & 24 & 2 & $<$0.001 & $<$0.001 & different \\
Claude Opus 4.8 (max) & Gemini 3.1 Pro (High) & 25 & 1 & $<$0.001 & $<$0.001 & different \\
Claude Opus 4.8 (max) & Qwen 3.7 Plus & 26 & 2 & $<$0.001 & $<$0.001 & different \\
Claude Opus 4.8 (max) & DeepSeek V4 Pro & 28 & 1 & $<$0.001 & $<$0.001 & different \\
GPT-5.5 (medium) & Claude Opus 4.7 (high) & 12 & 9 & 0.664 & 1.000 & tie \\
GPT-5.5 (medium) & Claude Opus 4.8 (high) & 13 & 9 & 0.523 & 1.000 & tie \\
GPT-5.5 (medium) & Claude Sonnet 4.6 (high) & 19 & 3 & $<$0.001 & 0.028 & different \\
GPT-5.5 (medium) & Gemini 3.1 Pro (High) & 19 & 1 & $<$0.001 & $<$0.001 & different \\
GPT-5.5 (medium) & Qwen 3.7 Plus & 19 & 1 & $<$0.001 & $<$0.001 & different \\
GPT-5.5 (medium) & DeepSeek V4 Pro & 23 & 2 & $<$0.001 & $<$0.001 & different \\
Claude Opus 4.7 (high) & Claude Opus 4.8 (high) & 9 & 8 & 1.000 & 1.000 & tie \\
Claude Opus 4.7 (high) & Claude Sonnet 4.6 (high) & 18 & 5 & 0.011 & 0.254 & tie \\
Claude Opus 4.7 (high) & Gemini 3.1 Pro (High) & 19 & 4 & 0.003 & 0.070 & tie \\
Claude Opus 4.7 (high) & Qwen 3.7 Plus & 18 & 3 & 0.002 & 0.043 & different \\
Claude Opus 4.7 (high) & DeepSeek V4 Pro & 20 & 2 & $<$0.001 & 0.003 & different \\
Claude Opus 4.8 (high) & Claude Sonnet 4.6 (high) & 16 & 4 & 0.012 & 0.271 & tie \\
Claude Opus 4.8 (high) & Gemini 3.1 Pro (High) & 16 & 2 & 0.001 & 0.039 & different \\
Claude Opus 4.8 (high) & Qwen 3.7 Plus & 18 & 4 & 0.004 & 0.112 & tie \\
Claude Opus 4.8 (high) & DeepSeek V4 Pro & 19 & 2 & $<$0.001 & 0.006 & different \\
Claude Sonnet 4.6 (high) & Gemini 3.1 Pro (High) & 8 & 6 & 0.790 & 1.000 & tie \\
Claude Sonnet 4.6 (high) & Qwen 3.7 Plus & 6 & 4 & 0.754 & 1.000 & tie \\
Claude Sonnet 4.6 (high) & DeepSeek V4 Pro & 8 & 3 & 0.227 & 1.000 & tie \\
Gemini 3.1 Pro (High) & Qwen 3.7 Plus & 6 & 6 & 1.000 & 1.000 & tie \\
Gemini 3.1 Pro (High) & DeepSeek V4 Pro & 8 & 5 & 0.581 & 1.000 & tie \\
Qwen 3.7 Plus & DeepSeek V4 Pro & 6 & 3 & 0.508 & 1.000 & tie \\
\bottomrule
\end{tabular}
\end{table}

\clearpage
\section{Licensing and Availability}
\label{app:licensing}

\textbf{Licences.} The benchmark-original code and tests are MIT licensed. Material from the \numrepos{} upstream projects follows each project's licence (Table~\ref{tab:task-manifest}): Upstream material is not redistributed. Setup scripts fetch each project from its public repository at a pinned commit and apply our task changes; this covers MIT-licensed projects as well as those with restricted or unclear terms (one project without a licence file, one with an author-stated copyleft licence, Lyra-derived plugins whose source is all rights reserved, and Epic content under the Unreal Engine EULA). For 19 tasks, an Epic engine plugin must be copied from a local Unreal Engine installation, and their start states were not verified end to end.

\begin{table}[H]
\centering
\small
\renewcommand{\arraystretch}{1.1}
\caption{The \numtasks{} tasks by upstream project, with each project's licence. Upstream code and content are not redistributed: setup scripts fetch each project at a pinned commit and apply the task's changes (the release includes only one MIT-licensed map file from UE-LASAA). Benchmark-original code and tests are MIT licensed. $^b$TargetVector bundles ALSXT and Ricochet (VOIDWARE dual licence), ALS-Refactored, TTToolbox, DataConfig, SPUD, and Modular Features Extra Actions (MIT), Lyra-derived plugins (all rights reserved; Epic content under the UE EULA), and EOS Integration Kit (UE EULA).}
\label{tab:task-manifest}
\begin{tabularx}{\linewidth}{@{}>{\raggedright\arraybackslash}p{0.23\linewidth}r>{\raggedright\arraybackslash}p{0.25\linewidth}>{\raggedright\arraybackslash}X@{}}
\toprule
Project (upstream) & Tasks & Task IDs & Licence \\
\midrule
HordeTemplateV2Native & 7 & 0001--0007 & MIT \\
ActionRoguelike & 10 & 0008--0015, 0017--0018 & no upstream licence file \\
Bomber & 21 & 0019--0031, 0034--0039, 0042--0043 & MIT \\
TargetVector & 34 & 0045--0065, 0067--0072, 0074--0075, 0077--0081 & VOIDWARE MIT (with organizational exclusions); bundled plugins mixed$^b$ \\
GASShooter & 7 & 0082--0083, 0085--0089 & MIT (code); Epic sample content under the UE EULA \\
Eternal Crusade: Resurrection & 19 & 0090--0108 & no upstream licence file (author-stated copyleft, unnamed); Lyra-derived parts under the UE EULA \\
UE-LASAA & 2 & 0109--0110 & MIT \\
PBCharacterMovement & 3 & 0111--0113 & MIT \\
NanoGaussianSplatting & 7 & 0114--0120 & MIT \\
\midrule
Total & 110 & & \\
\bottomrule
\end{tabularx}
\end{table}

\textbf{Release.} The release comprises the grader, the frozen test suites, the task list with the pinned upstream commits, the per-submission verdicts, the mutants, and the task packages or setup scripts; agent trajectories are not released because of their size and third-party content.

\end{document}